\documentclass[sigconf]{acmart}
\usepackage[utf8]{inputenc}

\usepackage{rotating}
\usepackage{adjustbox}
\usepackage{pdfpages}
\usepackage{caption}
\usepackage{subcaption}
\usepackage{soul}
\usepackage{array}
\usepackage{graphicx}
\graphicspath{ {./figures/} }
\usepackage{wrapfig}

\usepackage{color, xcolor,colortbl}
\usepackage{tabularx}
\usepackage{multirow}
\usepackage[utf8]{inputenc}

\newcommand{\del}[1]{\textcolor[rgb]{.4,.4,.4}{}}

\newcommand{\pOne}{P1} 
\newcommand{\pTwo}{P2} 
\newcommand{\pThree}{P3}
\newcommand{\pFour}{P4}

\newcommand{\pSix}{P6} 
\newcommand{\pSeven}{P7} 

\newcommand{\pNine}{P9} 
\newcommand{\pTen}{P10} 

\newcommand{\pictoEnergy}{\raisebox{-.2em}{\includegraphics[height=1.15em]{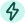}}}
\newcommand{\pictoData}{\raisebox{-.2em}{\includegraphics[height=1.15em]{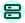}}}
\newcommand{\pictoDevices}{\raisebox{-.2em}{\includegraphics[height=1.15em]{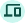}}}
\newcommand{\pictoLifetime}{\raisebox{-.2em}{\includegraphics[height=1.15em]{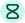}}}
\newcommand{\pictoUse}{\raisebox{-.2em}{\includegraphics[height=1.15em]{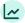}}}

\AtBeginDocument{%
  \providecommand\BibTeX{{%
    \normalfont B\kern-0.5em{\scshape i\kern-0.25em b}\kern-0.8em\TeX}}}

\setcopyright{none}
\acmDOI{https://doi.org/10.48550/arXiv.2507.19094}
\acmISBN{}

\acmConference[LIMITS '25]{11th Workshop on Computing within Limits}{June 26--27, 2025}{Online}
                  
\begin{document}

\title{Environmental (in)considerations in\\the Design of Smartphone Settings}

\author{Thomas Thibault}
\email{thomas@designersethiques.org}
\affiliation{%
  \institution{Designers Éthiques}
  \city{Paris}
  \country{France}
}

\author{Léa Mosesso}
\author{Camille Adam}
\author{Aurélien Tabard}
\affiliation{%
  \institution{LIRIS, Université Claude Bernard Lyon 1}
  \city{Lyon}
  \country{France}
}

\author{Anaëlle Beignon}
\author{Nolwenn Maudet}
\email{nmaudet@unistra.fr}
\affiliation{%
 \institution{Université de Strasbourg}
 \city{Strasbourg}
 \country{France}}

\renewcommand{\shortauthors}{Thibault, et al.}

\begin{abstract}
Designing for sufficiency is one of many approaches that could foster more moderate and sustainable digital practices. Based on the Sustainable Information and Communication Technologies (ICT) and Human-Computer Interaction (HCI) literature, we identify five environmental settings categories. However, our analysis of three mobile OS and nine representative applications shows an overall lack of environmental concerns in settings design, leading us to identify six pervasive anti-patterns. Environmental settings, where they exist, are set on the most intensive option by default. They are not presented as such, are not easily accessible, and offer little explanation of their impact. Instead, they encourage more intensive use. Based on these findings, we create a design workbook that explores design principles for environmental settings: presenting the environmental potential of settings; shifting to environmentally neutral states; previewing effects to encourage moderate use; rethinking defaults; facilitating settings access and; exploring more frugal settings. Building upon this workbook, we discuss how settings can tie individual behaviors to systemic factors.
\end{abstract}




\maketitle

\section{INTRODUCTION}
Despite impressive efficiency gains, the overall environmental impact of Information and Communication Technology (ICT) is steadily growing \cite{kara_comparison_2019}. More efficient infrastructures foster the development of applications that collect, store and transfer more data, and that are more computationally intensive. As these more intensive services become the norm, they provoke an increase in demand that leads to device renewal and the development of scaled-up infrastructures. Preist et al. frame this dynamic as part of a cornucopian paradigm \cite{preist_understanding_2016}. Tackling this challenge requires a wide breadth of complementary approaches ranging from regulatory changes, to changes in industry practices, or consumption practice~\cite{widdicks_breaking_2019}.

Among these approaches, Blevis highlights the role of design and HCI \cite{blevis_sustainable_2007}. Specifically, Widdicks et al. argue that designers should support users when they want to moderate their digital uses as this can align with their needs \cite{widdicks_breaking_2019,widdicks_escaping_2022}. As Preist et al. showed \cite{preist_evaluating_2019}, options offered to users could have a strong impact on emissions savings. When considering that 50\% of YouTube use is for music, if such streams were audio only, it would entail ``an emission reduction comparable to running a data center on renewable energy''
~\cite{preist_evaluating_2019}. Lite version of mainstream applications such as Instagram or Facebook Messenger also suggest space for designs aware of data, performance and connectivity limits. But such applications offer limited control over settings and are not available in many parts of the world. To go beyond these anecdotal examples, we examine which settings users have access to if they wish to control and moderate their mobile use and environmental impact.

More specifically, we are interested in understanding  (\textbf{RQ1}): \emph{to what extent smartphone settings support or prevent people from adopting, explicitly or not, environmentally aware practices?} Building on this first question, we further investigate  (\textbf{RQ2}): \emph{What design principles could support environmental settings?} i.e., settings that help users moderate their use when appropriate?

\section{RELATED WORK}
Our work builds upon the scholarship in Sustainable Interaction Design (SID) that investigates the role design could play in tackling ICT impact. The growing literature on Sustainability in Design \cite{blevis_sustainable_2007} can guide our attention to what settings should manage to be meaningful and impactful. Beyond efficiency in energy consumption, data transfer and storage, device lifespan and device proliferation, we also look at some socio-technical factors emphasized by Widdicks et al. on moderate use  \cite{widdicks_breaking_2019,widdicks_undesigning_2018}. 

\subsection{ICT Footprint and Moderating Use Strategies}
In computer science, environmental concerns have largely centered around \pictoEnergy \textbf{reducing energy consumption}. This has often been tackled through improvements in performance and through optimization~\cite{kurp_green_2008}. 
While performance issues on end-user devices can increase energy consumption \cite{noureddine_preliminary_2012}, poor performance become much more impactful in data-centers where computing power is concentrated. In data-centers, savings due to efficiency improvements happen at scale~\cite{lefevre_designing_2010}, with direct financial benefits which incentivizes performance gains, at the risk of rebound effects if \pictoUse \textbf{moderate uses} are not introduced alongside.

Networks and data transfer are another area of concern. The rapid growth in data transfer volumes, especially due to streaming, has been studied in various contexts. Preist et al. \cite{preist_evaluating_2019} show how various design interventions on YouTube design and default settings could significantly \pictoData \textbf{reduce data transfer and related emissions}, a finding later confirmed and extended by Suski et al. \cite{suski_all_2020}. Preist et al. \cite{preist_evaluating_2019} also note that high consumption can push for changes at the infrastructure level (e.g., the shift from 4G to 5G). Such large-scale deployment of new infrastructure has significant impacts.

Besides energy consumption of devices, data-centers and networks, Freitag et al. \cite{freitag_real_2021} estimate that device manufacturing accounts for about one third of ICT emissions, while a study of French ICT emissions assigns up to 50\% of emissions to manufacturing\footnote{Evaluation de l'impact environnemental du numérique en France. Ademe. 
Retrieved April. 28, 2025 from \url{https://librairie.ademe.fr/societe-et-politiques-publiques/7880-9522-evaluation-de-l-impact-environnemental-du-numerique-en-france.html}}.
\pictoLifetime \textbf{Extending the lifespan of devices} would lead to significant emissions reduction, and lessen stress on the environment. 
For example, Mosesso et al. \cite{mosesso_obsolescence_2023} showed that for smartphones, \pictoData \textbf{lack of storage} led to discarding otherwise working devices. 
Blevis \cite{blevis_sustainable_2007} also voiced concerns on device disposal, renewal and reuse, questioning the role of design in shortening or extending the lifespan of devices.

Another point of concern is digital lock-ins: it is becoming increasingly difficult to conduct basic activities without digital access~\cite{beignon_design_2021}, generating strong \pictoDevices \textbf{dependencies on devices} combined to the \pictoDevices \textbf{proliferation of devices} (handhelds, wearables, IoT) \cite{bates_exploring_2015,widdicks_streaming_2019}, sometimes needed to authenticate oneself, or perform administrative duties (some requiring access to a desktop computer, others to a smartphone). The dependencies also encompass an increasing expectation of permanent connectivity~\cite{lord_demand_2015,vinet_everyday_2023}. This relates to questions of \pictoUse \textbf{moderate use} (i.e. reduced use) \cite{widdicks_breaking_2019,hill_mapping_2020}, \pictoUse \textbf{control over use intensity} and designs that seek to capture attention or increase engagement \cite{monge_roffarello_towards_2022}.

\subsection{Settings and Software Customization}
We are interested in understanding how design can contribute to supporting people who would like to moderate their uses. This comes in complement from other approaches developed in ICT for sustainability in which developers make decisions on the level of consumption, energy expense, etc. Both approaches are useful, we investigate here how to support users in adapting their patterns of use and consumption practices to their own needs and contexts through dedicated settings. This leads to a second thread of research relevant to our goal: the study of software customization.

In the 1990’s, in the context of the ``feature war'', software vendors competed with each other over the number of features offered, leading to the development of software bloat \cite{mcgrenere_are_2000}. 
Coping with software bloat involved offering more control and customization possibilities for users to avoid overloading users. However, early research highlighted the many barriers that inhibit users’ customization practices, such as lack of time or knowledge \cite{mackay_triggers_1991}.

Over the past 20 years, the digital landscape has profoundly changed. Customization options and settings were plentiful on PC but they are less common in mobile applications. This evolution can be explained by two dynamics: with the increasing popularity of the smartphone has come a shift towards a ``one task equals one app'' model. Tchernavskij~\cite{tchernavskij_decomposing_2017} described these siloed applications as problematic for users: ``since it is impossible to meet the needs of every member of some particular community of practice, apps end up being designed from a one-size-fits-all approach''. The second factor is a shift towards algorithm-initiated customization which is thought of by its proponents as the solution to the overwhelming possibilities of customization and some of its limitations such as lack of time. However, research regularly shows that users tend to prefer mixed-initiative customization over more automatic adaptions~\cite{mcgrenere_evaluation_2002,sundar_personalization_2010}. Recently, possibilities offered through customization have been explored in the context of personal data curation~\cite{vitale_data_2020} for example, but to our knowledge no research has been conducted to investigate how much customization in mobile applications supports or prevents people from adopting more moderate and sustainable digital practices.
\section{ANALYZING ENVIRONMENTAL SETTINGS}
To answer our first Research Question, \emph{how much smartphone settings support or prevent people from adopting, explicitly or not, environmentally aware practices}, we conducted a study structured in three stages. For the first two stages, we adapted Hasinoff and Riven's feature analysis method~\cite{hasinoff_feature_2021}.
Originally designed to explore the relationship between app design and ideology, here we specifically analyze the relationship between settings design and the environmental impact of digital uses. We therefore chose to divide our analysis into the following three steps: 
 (1) We first analyzed whether environmental settings exist in mobile operating systems (OS) and applications. 
 (2) When they did, we analyzed how their current implementation supports (or not) users in understanding and using them. 
 To complement the results of this first analysis,
 (3) we organized two participatory workshops (18 participants in total) to get a better understanding of the perception and use of environmental settings.

\subsection{Settings Analysis Method}
\subsubsection{App and OS Selection}
\begin{figure*}[h]
         \vspace{-.3cm}
        \centering
         \includegraphics[width=1\textwidth]{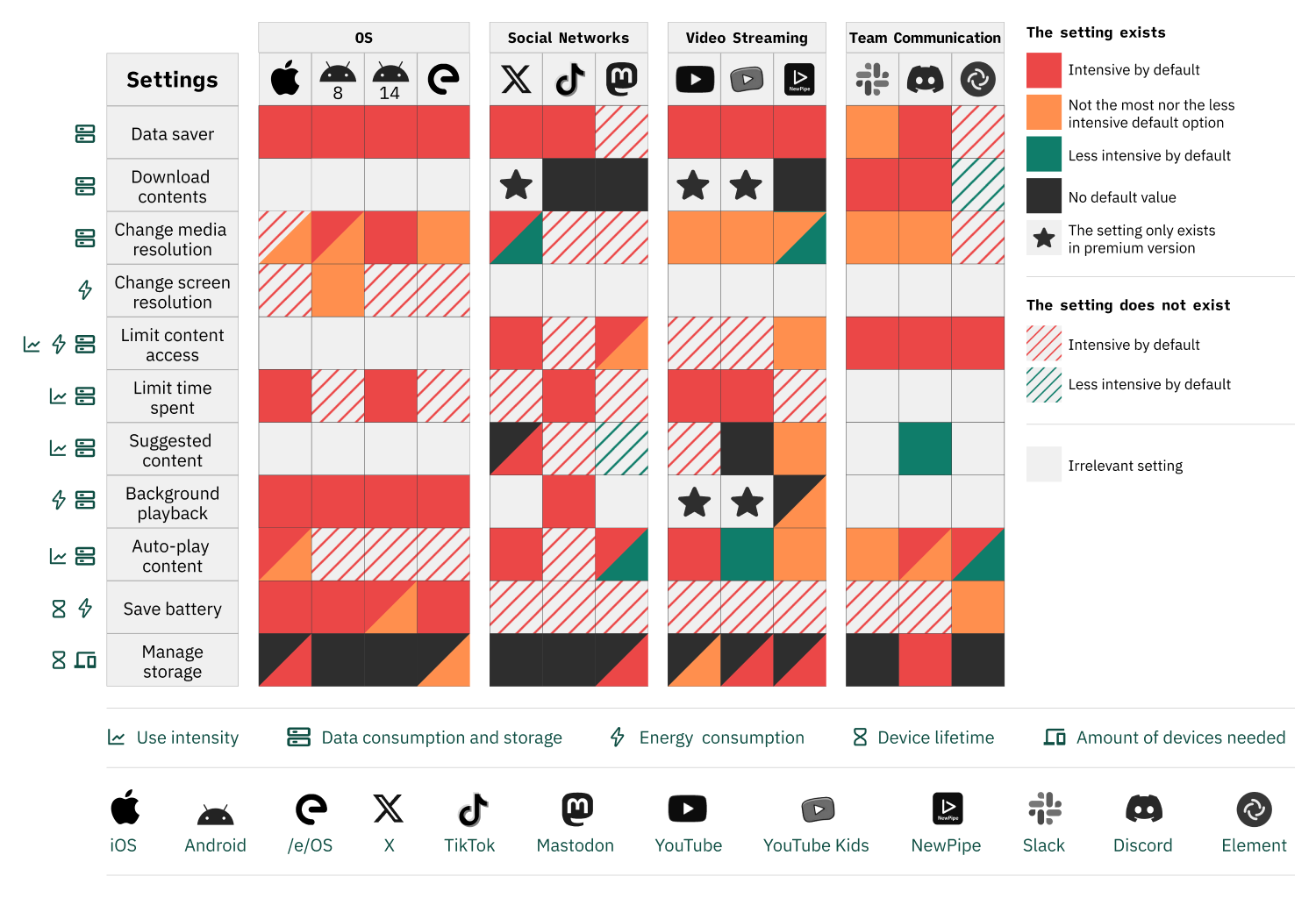}
         \vspace{-.8cm}
         \caption{↑ Summary table displaying the identified environmental settings and their default states}
         \label{fig:table}
         \vspace{-.1cm}
\end{figure*}

Because there are settings in almost every application, we first selected a representative and diverse sample of applications to analyze. We chose to focus on three complementary categories of widely used applications on smartphones that generally provide many settings : social networks, video streaming and team communication. We selected two popular applications for each category as well as an open source alternative.
As settings exist both at the app and the OS level, we also included three operating systems in our corpus, the two mainstream ones (iOS and Android) as well as an open source alternative\footnote{iOS devices are kept much more up to date, so we picked a recent version at the time of analysis. Android devices on the other hand offer less updates so we used 2 different versions. We also chose an Open Source alternative presenting itself as long-lasting and ecological.}: 
iOS (version 15.8.3), Android (versions 8.0 and 14.0) and /e/OS, a free OS building upon Android and Lineage (version 2.3) [Fig. \ref{fig:table}].

\subsubsection{Environmental settings identification and analysis}

The first goal of the analysis was to identify what we call environmental settings, i.e. parameters that affect the various environmental impacts of digital technology.
We identify settings as `environmental' even when they are not explicitly presented as such. 

\newpage
From our literature review, we synthesized and defined five environmental settings categories:


\begin{itemize}
    \item[\pictoData] \textbf{Data consumption and storage}, e.g., limiting mobile data usage 
    \item[\pictoEnergy] \textbf{Energy consumption (on the device or on remote servers)}, e.g., brightness, always on sensors
    \item[\pictoUse] \textbf{Use intensity}, e.g., infinite scroll, autoplay
    \item[\pictoDevices] \textbf{Amount of devices needed}, e.g., two-factor authentication
    \item[\pictoLifetime] \textbf{Device lifetime}, e.g., battery health, assistive touch for damaged screens
\end{itemize}

\begin{figure*}[h]
    \centering
    \vspace{-.1cm}
    \includegraphics[width=1\textwidth]{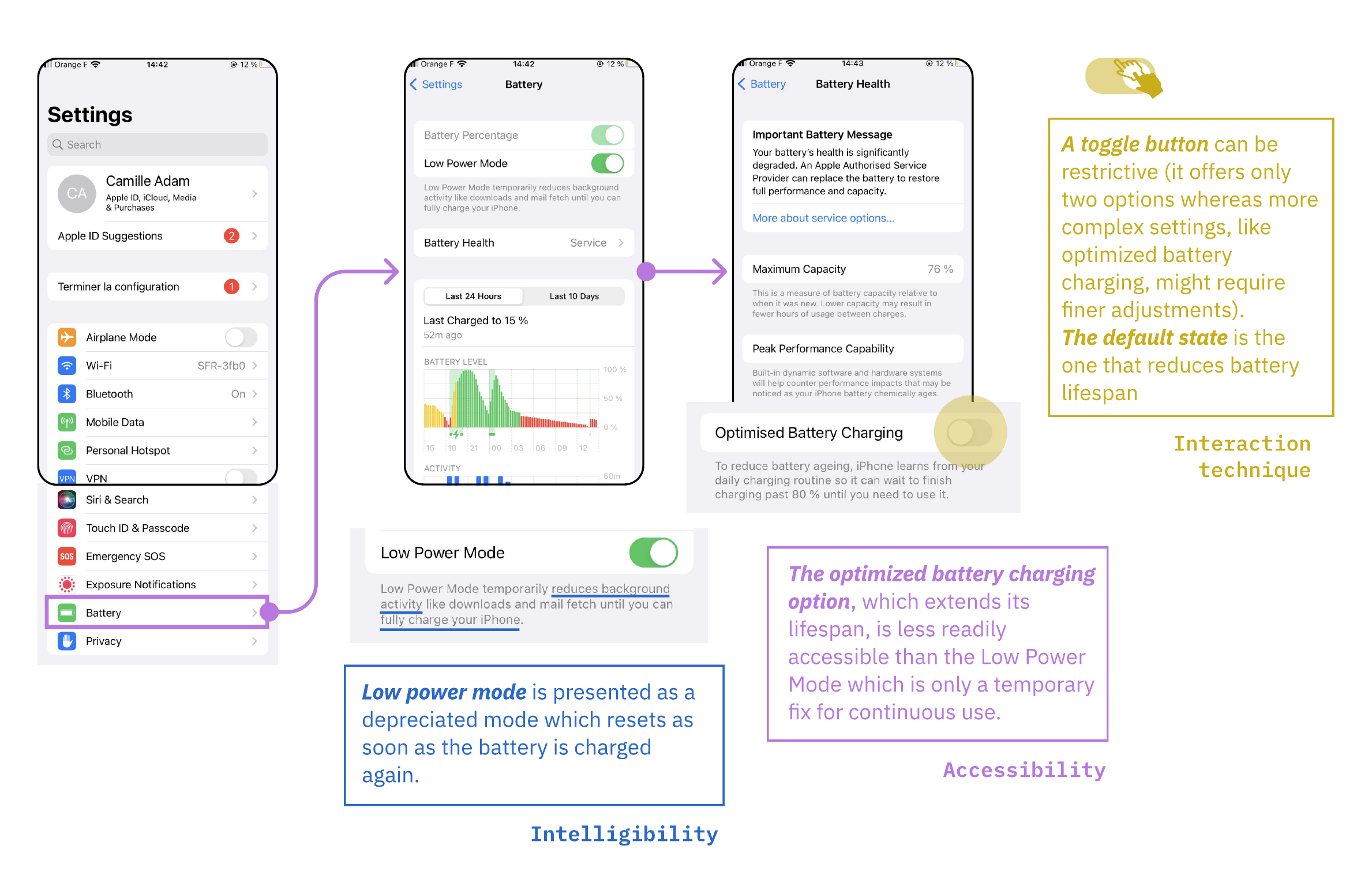}
    \caption{↑ Example of the ``save battery'' interface analysis in iOS}
    \label{fig:settingsanalysis}
\end{figure*}

After defining these environmental settings categories, we analyzed settings using the following method:
(1) We went through each application and OS of our corpus, searching manually in all the menus for environmental settings. We assigned the settings to a given category and we also recorded settings default value and compared it with the other available options to determine if it is intensive by default. In our analysis, the intensity of setting is therefore not assessed on absolute terms but rather in relative terms: is the default option relatively more intensive or less intensive than the other proposed options. We compiled our results in a summary table [Fig. \ref{fig:table}].
(2) As Hasinoff and Riven explain \cite{hasinoff_feature_2021}: ``the outcomes that developers expect their app to produce are not simply or directly determined by an app’s features [...] [the] mediating role of affordances is vital for feature analysis." In our case, we chose to analyze each environmental setting in terms of accessibility, intelligibility and interaction qualities to understand how users might be able to understand and/or use them. We summarized the access path to each setting in color-coded navigation maps. 
[Fig. \ref{fig:settingsanalysis}]
After assessing their accessibility (understood here as menu depth \cite{bailly_visual_2017}), we analyzed their intelligibility, i.e., graphic and iconographic design, UX copy (naming and explanation) as well as the interaction techniques used (toggle button, slider, etc.) for each setting category. From our analysis, we identified what we call six anti-patterns that show the current inconsideration in the design of smartphone settings. Anti-patterns have been defined as "a commonly occurring solution to a problem that generates decidedly negative consequences” \cite{brown1998antipatterns}. In HCI, Widdicks et al. have shown that anti-patterns do not necessarily occur because of intentional design decisions \cite{widdicks2020backfiring}. Nevertheless, identifying these patterns is a prerequisite to help designers avoid them.
As a first verification of these anti-patterns, we looked beyond our original corpus and identified at least one other example for each anti-pattern in other applications and/or websites.

\subsection{Understanding Perception and Use Through Participatory Workshops}

After conducting our analysis, we chose to follow Hasinoff and Bivens’ advice to complement our initial results by studying users \cite{hasinoff_feature_2021} through two participatory workshops. Our goal was to better understand the perception and use of environmental settings in practice. The first workshop took place at a local environmental association with 6 participants and lasted for 2 hours.
The second one took place in a library with 12 participants and lasted 1 hour and a half. To recruit participants, we advertised the workshops through the association and the library communication channels. Participants were not compensated, they participated because they wanted to know more about settings and their challenges. In exchange for their participation, we provided them with support and advice to manage their settings.



Both workshops followed identical schedules: (1) As an introduction to the workshop and without explaining what we identified as environmental setting, we asked participants about their perception of the environmental aspects of settings. We were especially interested in their perception of ecological leeway in smartphone interfaces and what ecological effects they thought different settings could have. 
Related to RQ1, our goal was to see if participants could identify which settings had an environmental impact. 

(2) After introducing the five environmental settings categories we identified, we asked them whether they could think of existing settings for each category. We noted each setting they mentioned, and then asked everyone if they had been using some of them. Our goal was to see if participants already used some environmental settings (RQ1). 

(3) Lastly, we presented them 11 settings we had selected across four of the five environmental settings categories. We provided them with a cheat sheet for each setting, featuring an explanation of their effects and how to locate them in iOS and Android. During the remaining time of the workshop, participants were invited to test at least one of these environmental settings on their phones and we accompanied them when they had questions or could not locate the setting on their applications and OS. We took notes of participants who voiced concerns about the consequences of activating or modifying some of the settings; or had issues when trying to locate or use them. Our goal was to identify the limits of existing settings from which we could build better design principles (RQ2).

The first two authors conducted the workshops. They both took notes during the workshops and met afterwards to ensure they had recorded all relevant comments and actions they witnessed. We then used them to cross-check our feature analysis. 
When reporting these findings, we chose not to provide quantitative metrics of how many participants made a given statement when discussing settings perception. As qualitative theorists have argued, ``numbers can lead to the inference (by either the researcher or the audience) of greater generality for the conclusions than is justified'' \cite{maxwell_using_2010}.


\section{RESULTS}
Both the settings and participatory workshops analyses show that there is a pervasive lack of environmental concern in settings design. Through our analysis, we could identify and define 6 anti-patterns in the design of settings on smartphones. We first present anti-patterns that exist at the individual settings level (4.1 to 4.4) before turning to larger-scale anti-patterns (4.5 and 4.6).

\subsection{Ecologically Relevant Settings Are Not Presented as Such}
While some existing settings can have an ecological impact, they are rarely designed or presented through this lens. For example, video quality can have a significant impact in terms of data consumption but the wording only highlights its effect on visual quality. 
Media quality settings are generally not included in energy or data settings categories. For example, video upload quality in Discord is classified under the Chat category, and Video autoplay in Twitter/X is classified in the Accessibility category.
Settings are instead often presented from a performance and consumerist point of view. For example, according to the wording, ``data saver mode'' exists primarily to save money. 


\begin{figure}[htp]
    \begin{subfigure}{0.45\textwidth}
        \centering
         \includegraphics[height=4cm]{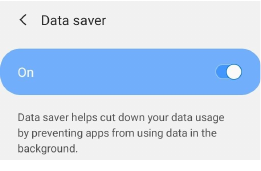}
         \vspace{-.2cm}
         \caption{↑ In Android, the Data saver mode provides no information on how much data is saved}
         \label{fig:datasaver}
     \end{subfigure}
    \bigskip
    \\
    \begin{subfigure}{0.45\textwidth}
        \centering
         \includegraphics[height=4cm]{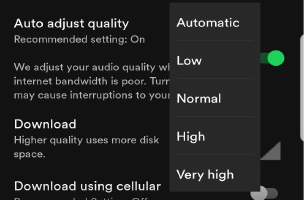}
         \caption{↑ In Spotify, the difference between “normal” to “very high” quality options is almost imperceptible}
         \label{fig:adjustquality}
    \end{subfigure}
    \vspace{-.2cm}
    \caption{Examples of settings that provide no explanation of their environmental impact}
\end{figure}

During our participatory workshops, participants were often surprised to discover that settings can have an ecological impact and had a hard time identifying some when prompted. For example, \pOne~ knew how to desynchronize their photos from the cloud, but they kept them synchronized, because they did not know that automatically uploading photos to the cloud eventually had environmental effects by increasing data consumption and storage as well as fostering the growth of digital infrastructures \cite{widdicks_demand_2017}.

\subsection{Settings Provide Little Understanding of Their Effects and Environmental Impacts}
Settings enabling explicit limitations, such as data saving mode, provide little details about their concrete effects (how much data or battery is saved and how) [Fig. \ref{fig:datasaver}]. 
Applications also offer very little insights on the effects of different options, e.g., comparison between quality levels beyond vague terms such as ``low'' or ``high''. And such designs do not take into account human perception capabilities [Fig. \ref{fig:adjustquality}], although for instance most people do not perceive the difference between HD and UHD (4K) video quality \cite{kara_comparison_2019}. Moreover, many applications (including YouTube at the time of our analysis) let users download high quality audio or video even when the device hardware does not support such high resolutions. 

During our participatory workshops, we also found that participants did not understand the environmental impacts of settings. \pTen~ did not know that using Wi-Fi was better than 4G (on average) in terms of carbon emissions \cite{huang_close_2012} and learned it during the workshop. They then chose to configure settings to better control their use of 4G: setting a data limit and adapting video quality to the context (Wi-Fi or 4G).

\subsection{When a More Frugal Option Does Exist, It Is Presented as a Lesser Option}
\begin{figure}[htp]
\centering
     \begin{subfigure}{0.48\textwidth}
         \centering
         \includegraphics[width=.78\textwidth]{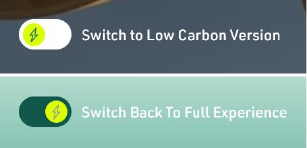}
        \caption{↑ Carbon button - cop28.com}
         \label{fig:cop28}
     \end{subfigure}
     \bigskip
     \\
     \begin{subfigure}{0.48\textwidth}
         \centering
         \includegraphics[width=.78\textwidth]{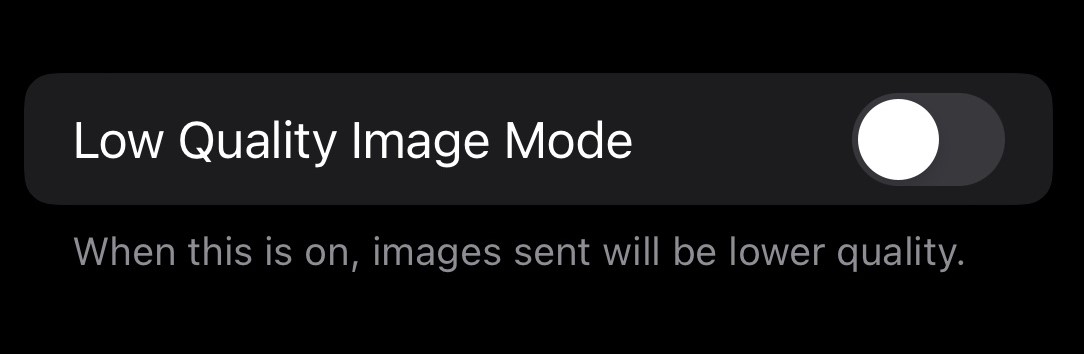}
         \caption{↑ Low Quality Image Mode in iMessage Settings}
         \label{fig:qualitymessages}
     \end{subfigure}
     \vspace{-.5cm}
     \caption{Examples of more frugal options presented as lesser options}
\end{figure}

As evidenced by the vocabulary used to describe settings effects, we found that most moderate options are presented as depreciated ones. Outside of our corpus for example, the COP28 website is emblematic of this issue [Fig. \ref{fig:cop28}]. On the main page, a toggle greets users with a ``switch to Low Carbon Version'' label. When activated, it says ``Switch Back To Full Experience''. The wording of the setting implies that the low carbon version provides an incomplete experience, even though a faster loading time could be perceived as a better experience. 

To fully integrate an ecological perspective is to think about compromise, an arbitration between performance and use comfort on the one hand, and ecological impact on the other hand. For example, iOS has a setting to lower image quality sent in messages. The settings states that ``when this is on, images sent will be lower quality''. But there is no mention that reducing the quality can save data, either from an economic or ecological point of view [Fig. \ref{fig:qualitymessages}].
Labeling settings with words such as ``high-quality'' versus ``low-quality'' implies that one option is better than the other. Therefore, users only see the negative aspects of the most frugal option.

\subsection{Default Choices Are the Most Intensive Ones}
Even when settings exist, designers have to set default values. These default choices are especially important because many users will not take the time to explore and configure their settings. One choice repeated on millions or billions of devices can have an important impact on the digital ecosystem. 

\begin{figure}[htp]
    \begin{subfigure}{0.4\textwidth}
        \centering
         \includegraphics[width=\textwidth, trim={0 .2cm 0 7.2cm},clip]{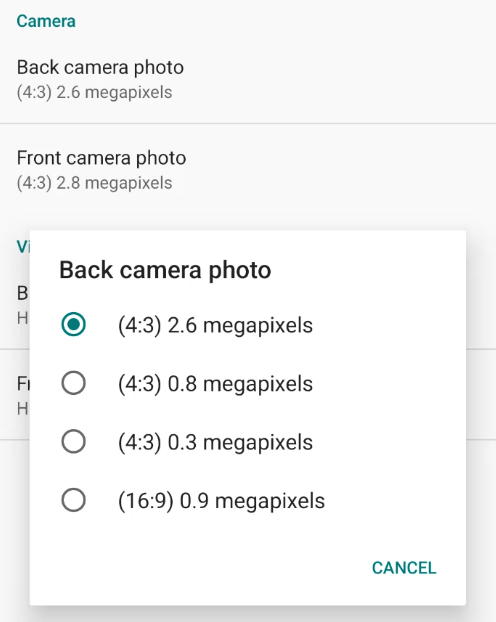}
         \caption{↑ Back camera photo size in Android is the highest quality by default
         }
         \label{fig:picturesquality}
     \end{subfigure}
    \bigskip
    \\
    \begin{subfigure}{0.48\textwidth}
        \centering
         \includegraphics[width=\textwidth]{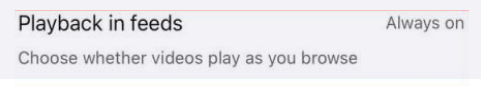}
         \caption{↑ In YouTube, video thumbnails automatically start playing when hovered, downloading a large amount of data}
         \label{fig:playbackinfeeds}
    \end{subfigure}
    \vspace{-.5cm}
    \caption{Examples of intensive default settings}
\end{figure}

Our analysis shows that the default option is very often the most intensive one, i.e., the one that consumes the most energy or data [Fig. \ref{fig:picturesquality}, \ref{fig:playbackinfeeds}]. In our corpus, this is the case for 61 out of the 96 settings we analyzed. 
These choices can be explained by designers' concern to show the maximum of an application, technology, hardware or content. Regarding photo resolution for example, in Android the default resolution is always the highest even if perceived photo quality depends less on the number of pixels and more on the image processing algorithms used to transform the camera sensor data into an image.

During the second participatory workshop, \pSeven~ realized that their battery saver option in Android was off by default. When activating it, the most intensive option was selected by default and \pSeven~ had to manually select the most frugal option.

Overall, when looked at individually, we found that most designs of settings do not take into account their environmental impact. Our analysis also identified issues at a larger and more systemic level.

\subsection{Environmental Settings Are Scattered and Hard to Locate}
Environmental settings are located inconsistently across OS and applications.
Users who have learned where a given setting is located (for example data saving mode) in one application cannot reuse this knowledge in other applications and must always rediscover it. This prevents users from efficiently customizing their mobile phones and further reinforces the impact of default choices.

Even for the same service, equivalent settings can be located in different menu hierarchies in the application and on the website. Equivalent settings are often named and presented differently in different applications, adding to the general confusion.
For example, in the YouTube application, the ``Playback in feeds'' setting is located in the application's general settings, but on the YouTube website, the same setting is located in the ``Playback and performance'' section, under the name ``Video previews''.
Moreover, many similar settings need to be individually set in each application instead of being located at the OS level.

A majority of participants in our workshops struggled to locate specific settings in their phones. We had anticipated this issue and were hoping to guide them using dedicated cheat sheets showing the generic path to find each setting. However, we found out during the workshops that paths significantly differ from one phone model to another and that some settings do not even exist on certain devices. 
We found out that participants were often confused by the vocabulary used to describe settings categories. For example, \pTwo~ was not able to find how to reduce media quality in Twitter/X because that setting was located in ``Accessibility, display and languages''.  \pSix~ got lost and ended up in the video-format setting menu when trying to find the setting that controlled video ``size'' in NewPipe, because they got confused with the technical terminology used.

In the settings hierarchy and overall organization, ecologically relevant settings are surrounded by multiple settings that are not relevant from an ecological perspective. This reduces the intelligibility of settings and inhibits practices gearing towards use moderation through better user control.

\subsection{There Are Fewer Settings in Mainstream Applications} 
When comparing proprietary applications to free and open source software (FOSS) ones, we found out that the former generally offer fewer settings. NewPipe, a free software alternative to YouTube, offers a lot more settings than YouTube. Mastodon, a free software alternative to Twitter/X, offers to delete posts after a period of time. This relates to its business model not based on data collection, but on decentralized instances carried by smaller communities, who must take care of management and hosting costs. Mastodon users and instance administrators therefore have less incentive to store old content.
Also, Mastodon is accessible from various client applications, each offering its own additional settings. Using any of these applications already offers a wider choice than the Twitter/X application alone.

During our workshops, we found that participants were interested in installing FOSS applications. Indeed, half of participants from the first workshop and one third of participants from the second workshop chose to install NewPipe, arguing that they would be able to use specific settings that they did not have access to in YouTube. These included environmental settings such as downloading audio only, but also settings that met other needs such as limiting one's incentive to binge-watch videos.

Open-source applications settings generally offer more control over the data and overall use of the service. A lot of settings in proprietary applications are focused on privacy and security, whereas open-source applications tend to emphasize privacy and security by default and do not need these settings.
We also found that open-source applications (Mastodon, NewPipe, Element) provided default values that were less intensive, compared to the mainstream proprietary applications we studied.
For example, by default, NewPipe does not provide the 2K and 4K options on the video player. If they want to have these options, users first need to enable them in their settings. NewPipe developers justify their choice in the setting menu by explaining that ``only some devices can play 2K/4K videos". Similarly, by default, Mastodon does not autoplay videos and automatic reading of GIFs is disabled.

\vspace{-.1cm}
\section{DESIGNING ENVIRONMENTAL SETTINGS}
\textbf{Approach and Objectives}
Our analysis of smartphone settings revealed how little care for environmental impacts there is in their design. We now explore what settings could look like if they were to foster more sustainable digital practices.
Building upon our analysis and to answer the first five anti-patterns in settings design, we identified five design principles: 
    \vspace{-.1cm}

\begin{enumerate}
    \item present the environmental potential of settings, \item previewing settings effects to encourage moderate use, 
    \item make frugal options desirable, 
    \item rethink default options, and 
    \item facilitate faster and more frequent access to settings. 
\end{enumerate}

To illustrate and test these principles, we used them as prompts for creating a design workbook \cite{gaver_making_2011}, i.e., a collection of design proposals that explore alternative settings designs to respond to the issues highlighted in our analysis. As Gaver explained, the objective of workbooks is not to design ``the right thing'' but rather to open, scope and map a vast underexplored design space. Indeed, ``the power of design workbooks is in creating a much larger landscape for exploring [...] concerns by exploiting the combinatorial explosion of similarities and differences among many such proposals''. 
We sought to develop both concrete and specific ideas that could be easily implemented in the current state of technology, and others that are more speculative in nature so as to inspire designers to go beyond traditional and currently expected ways of thinking and designing settings. 
After two collective brainstorming sessions to elicit ideas following our five design principles, we realized that we had also explored ideas that went beyond the aforementioned design principles. After reviewing them, we chose to group them by affinity into three additional design principles (section 5.6): 
    \vspace{-.1cm}

\begin{enumerate}
    \setcounter{enumi}{5}
    \item provide settings for aging phones 
    \item provide settings for voluntary limitation
    \item provide collective settings
\end{enumerate}
We then created mock-ups to document and communicate the ideas. These visuals are included as supplementary materials \footnote{\textbf{Environmental Settings Design Workbook}: \url{https://limitesnumeriques.fr/media/pages/travaux-productions/design-ecolo-parametres/ff9916ba8d-1745592583/supplementary-materials.pdf}}

\subsection{Present the Environmental Potential of Settings}
Settings should better reflect the materiality of digital technologies and their various environmental impacts by using visuals and words that communicate it. 
Simple changes such as changing settings labels can already help users understand the impact of their choices.
\begin{itemize}\setlength{\itemsep}{-.5mm}
\item \emph{Autoplay next video} (YouTube) → \emph{Auto-download next video}
\item \emph{Low-quality video} → \emph{Low-impact video}
\end{itemize}

Because metaphors are powerful design tools\cite{blackwell_reification_2006}, we also suggest to move away from metaphors like the ``cloud'' or icons that hide the materiality of digital infrastructure or their potential impacts. ``Airplane mode'' could be called a ``Disconnect mode'' or ``Battery-saving network mode''.
During our workshops, \pNine~ noted that vocabulary like ``the cloud'' obfuscates meaning, thinking it was deliberately designed so by corporations.  

\subsection{Preview Settings Effects to Encourage Moderate Use}
\begin{figure}
    \captionsetup{justification=centering}
    \begin{subfigure}[b]{.45\columnwidth}
    \centering
         \includegraphics[width=1\columnwidth]{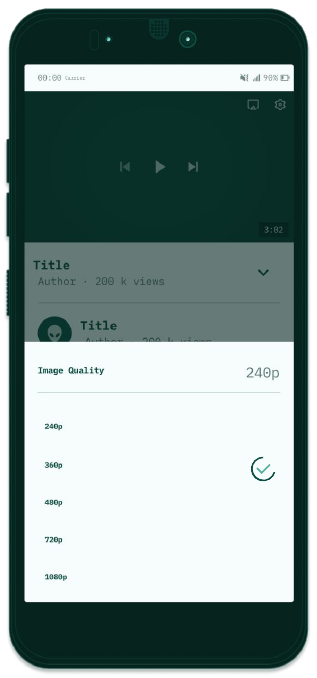}
         \caption{↑ Choosing a higher image quality takes longer than choosing a lighter one 
         }
         \label{fig:feelingweight}
    \end{subfigure}
    \hfill
    \begin{subfigure}[b]{.45\columnwidth}
        \centering
        \includegraphics[width=1\columnwidth]{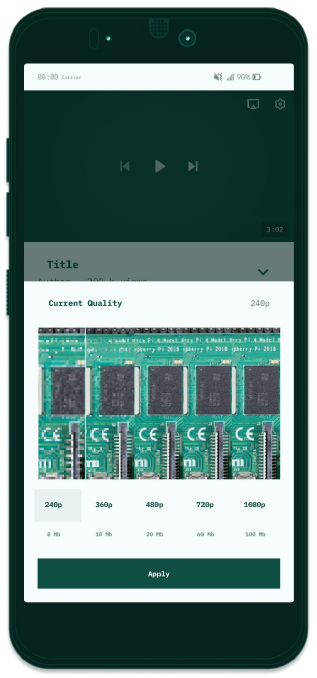}
        \centering
        \caption{↑ Each quality option is previewed\\ to facilitate direct comparison}
        \label{fig:comparing}
     \end{subfigure}
     \caption{Previewing Settings Effects}
\end{figure}

Settings should display their potential in terms of energy or data savings, moderating use, device lifespan, etc., either quantitatively, qualitatively or comparatively, through visual or interactive means. When possible, they should explain how battery endurance will be improved, or how much data will be used to provide a given image quality. They should also help users understand that their choices are generally compromises (e.g., more compute performance leads to shorter battery life). 
For example, selecting a more demanding option could require more effort or feel slower or heavier than choosing a more frugal option. In this way, users could perceive that an option consumes more than another [Fig. \ref{fig:feelingweight}]. 
But settings could also come with visual examples that help users see if they can really perceive the quality difference instead of, e.g., choosing a video or audio quality based on numbers
[Fig. \ref{fig:comparing}].

\subsection{Make Frugal Options Desirable}
\begin{figure}[htp]
    \includegraphics[width=.43\textwidth]{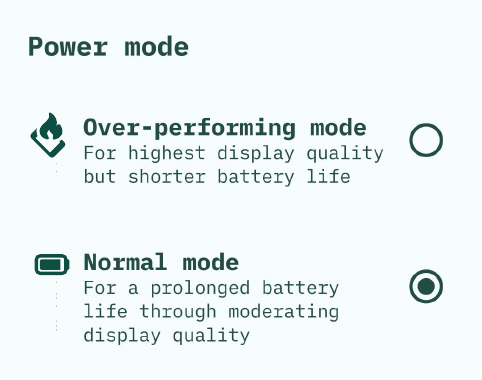}
    \vspace{-.2cm}
    \caption{↑ The low-quality mode becomes the normal mode}
    \label{fig:neutral}
\end{figure}

We also need to change what is considered a neutral state. When presenting different options, the baseline should be the most sustainable option and other options should be presented in comparison to this baseline. For example, we could choose to present low quality modes as normal modes and higher quality ones as over-performing modes [Fig. \ref{fig:neutral}].

\subsection{Rethink Default Options}
\begin{figure*}[htp]
    \centering
    \hfill
    \begin{subfigure}{0.3\textwidth}
        \centering
         \includegraphics[width=.8\textwidth]{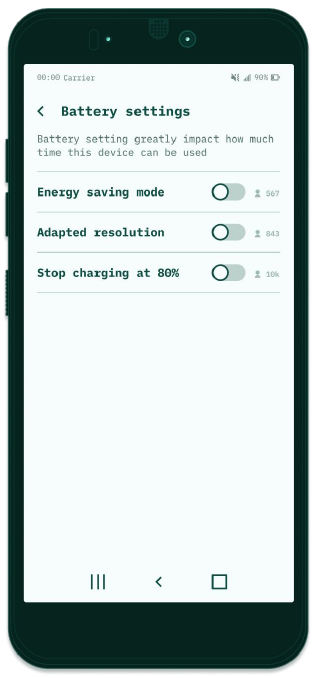}
         \caption{↑ The number of users modifications for each setting is visible in the Settings app\\}
         \label{fig:collectivecount}
    \end{subfigure}
    \hfill
    \begin{subfigure}{0.3\textwidth}
        \centering
         \includegraphics[width=.8\textwidth]{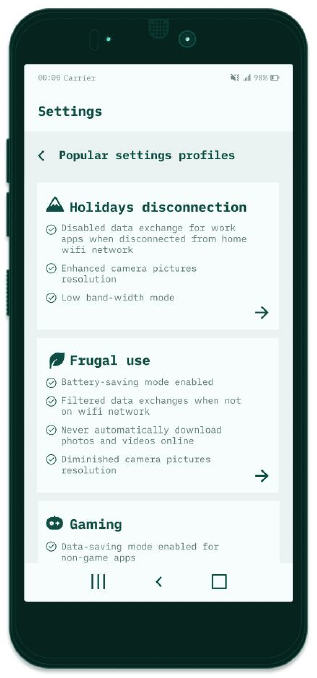}
         \caption{↑ Profiles of settings are suggested to users to help them with selecting options according to their uses}
         \label{fig:collectivesuggestions}
     \end{subfigure}
     \hfill
    \begin{subfigure}{0.3\textwidth}
        \centering
        \includegraphics[width=.8\textwidth]{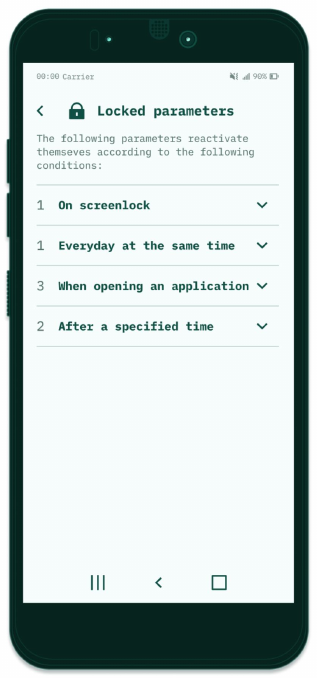}
        \caption{↑ Users can define a default photo quality which is restored each time they unlock their screen - \href{https://drive.google.com/file/d/1HbnbNBH9npyWClFHSCGOmzXjvUZ0yRuI/view?usp=sharing}{Animation Link}}
         \label{fig:default}
     \end{subfigure}
     \caption{Rethinking Default Options}
\end{figure*}

We think that one of the most impactful decision regarding environmental settings would be for designers to define defaults as the most frugal option. But we can also help users define their own default options and provide means of restoring frugal default options after periods of intensive use.
One way would be to provide ``community defaults": users might feel uncomfortable changing the official default options; we could help them by showing which options are the most commonly used ones [Fig. \ref{fig:collectivecount}]. When downloading an application, designers could provide sets of default settings: community presets, or presets related to a type of use [Fig. \ref{fig:collectivesuggestions}].
Another idea is to automatically reset to frugal options: we saw that users activate options such as Bluetooth or 4G when they need them, but can very easily forget to deactivate them. We propose an ``automatic moderate reset'', after a set time or on application reboot, so that the mobile always defaults to the most frugal option [Fig. \ref{fig:default}].

\subsection{Facilitate Faster, More Frequent, and More Legible Access to Settings}
We argue that facilitating access to and use of environmental settings is one way of helping users adopt more environmentally aware practices by enabling them to adapt their consumption to their specific needs and situations of use, instead of having settings set to the maximum performance by default.
Letting users rename settings themselves could help them relocate settings and appropriate them more easily. ``Low screen resolution'' could be renamed ``Sparing battery mode'' as it reduces the phone's computing needs. This idea could improve how settings are signified to people with little customization experience or interest. For instance, a person with more experience could rename features for a relative, a colleague, or a friend who has difficulty navigating and managing settings. This could contribute to democratize customization 

\begin{figure*}
    \centering
    \hfill
        \begin{subfigure}{0.3\textwidth}
         \includegraphics[width=\textwidth]{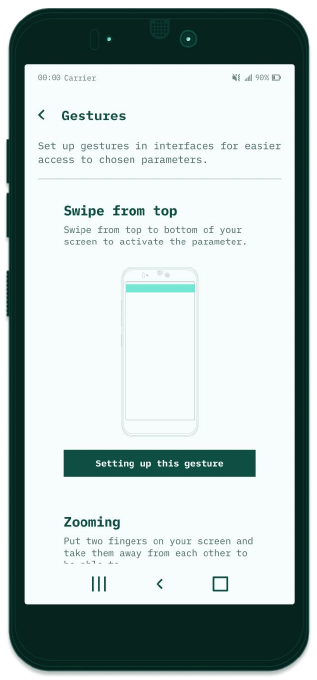}
         \caption{↑ Using a gesture (e.g., sliding with 3 fingers or a hand) to decrease and increase quality - \href{https://drive.google.com/file/d/1NXUZMU-FSNWGqUCpeTOZmMEhPUKyuiO9/view?usp=sharing}{Animation Link}}
         \label{fig:gesture}
     \end{subfigure}
    \hfill
    \begin{subfigure}{0.3\textwidth}
         \includegraphics[width=\textwidth]{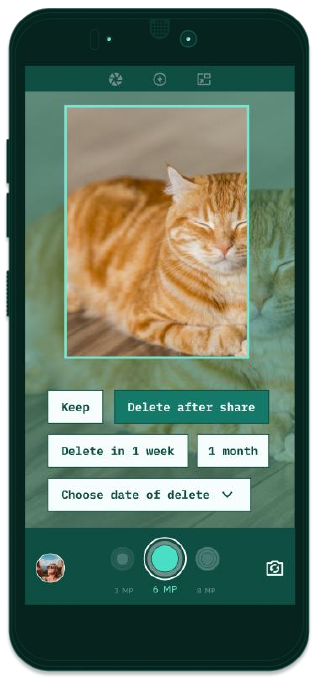}
         \caption{↑ Users can choose the expiry date when saving a photo\\
         }
         \label{fig:savingphoto}
     \end{subfigure}
     \hfill
        \begin{subfigure}{0.3\textwidth}
         \includegraphics[width=\textwidth]{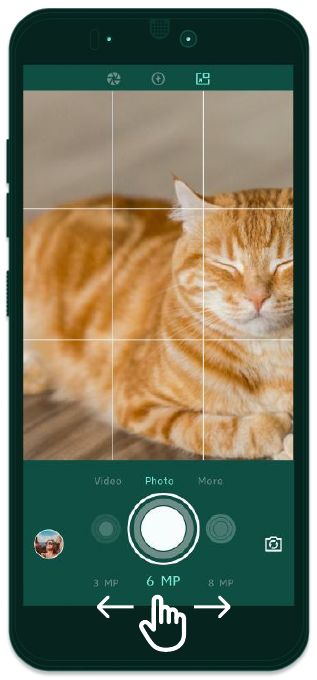}
         \caption{↑ Users can select photo quality by dragging the button before shooting - \href{https://drive.google.com/file/d/1r92N4LZ1p3UBqTmdxnRtEcl8W4m7FGdN/view?usp=sharing}{Animation Link}
         }
         \label{fig:dragging}
        \end{subfigure}
     \hfill
     \caption{Facilitating access to settings}
\end{figure*}

As we have seen in our analysis, environmental settings are currently distributed and buried in complex menu hierarchies. Shortcuts could help users choose to activate/deactivate settings more easily: through a combination of buttons, through defined gestures on the screen [Fig. \ref{fig:gesture}], through physical sensors, etc. For example, if the phone is held upside down, only the sound of a video would be downloaded. Internal thresholds such as saturated storage, data consumption limits or even external ones such as an excessively carbon-intensive electricity mix could also be used as settings triggers. 
We also think that rather than leaving settings in a separate space and with menus detached from their impact, settings could be better integrated within applications UI. For example, a camera application could provide contextual interactions to set the expiry date and size of a photo when taking the photo [Fig. \ref{fig:savingphoto}, \ref{fig:dragging}].

\subsection{Provide Settings for Aging Phones}
While the main way to reduce the impact of a smartphone is to extend its lifespan, none of the settings in our analysis focused explicitly on adapting a phone’s features to its gradual deterioration. To our knowledge, the Assistive Touch function in iOS is the only one that lets users bypass button breakage by recreating on-screen buttons. In the same way, we could extend this feature to bypass other physical failures. For example, we could enable users to move or resize essential interaction elements so that they no longer lie underneath a broken screen part; Schaub et al. developed some of these ideas further \cite{schaub_broken_2014}. 


\subsection{Provide Settings for voluntary limitation}
Today's digital technologies are designed to be ever more seamless~\cite{lorusso_user_2021}.
This fluidity is indifferent to the quantity of resources mobilized, preventing users from being aware of their data and energy consumption. 
Settings could enable users to create voluntary limitation or friction, like an expiry date to contents and applications in order to adapt to storage saturation. For example, users could choose the fate of an application when installing it: ``pop-up'' for applications that delete themselves after one month of non-use (useful for applications downloaded while traveling, etc.), or ``dispensable'' for applications that automatically delete themselves when full storage is reached 

More broadly, in our workshops, several participants were interested in settings that could help them use their phone less, leading to less environmental impacts through decreasing data transfer and reducing battery discharge. \pOne~ and \pSix~, for example, wanted to put on the grayscale mode, and \pOne~ also suggested their phone could be ``less addictive'' during electricity consumption peaks.

\subsection{Provide Collective Settings}
During our participatory workshops, many participants explained that they dared testing settings because they were in groups and surrounded. 
\pOne, \pThree~ and \pFour~ said that they would not have changed their settings if they were alone. Thinking of environmental settings as a collective practice could help amplify their potential impact.
We could, for example, let users share individual settings or sets of settings to support other users who would like to reduce their data demand or extend their battery life.
Moreover, we could have have a function that allows an application to be fully configured remotely by someone else, for a certain period of time. We could also support community-defined settings through collective votes on the quality and expiry date of media in a messaging group, for example 





\section{Discussion, Limitations and Future Work -- Are Environmental Settings Worth the Effort?}

The settings we criticized, proposed, and reflected upon are to be set on individual devices. This is something that many scholars working on Sustainable HCI have warned about, emphasizing the importance of scale~\cite{dourish_hci_2010, bremer_have_2022}. Putting the responsibility of change on individuals can contribute to minimizing the cultural, political and systemic issues at play. 
In this discussion, we reflect on the systemic elements of settings, the balance of user-led micro-changes and macro-changes but also in-between options such as collective customization. We also discuss the risks of using settings for greenwashing. While our paper focused on smartphone settings as a case study, this discussion is relevant to a broader set of devices and contexts.

\subsection{Empirical Evidence of Environmental Improvements}
One limitation of this work is that we did not quantify the environmental impact of the settings studied, nor did we rank design ideas according to their estimated impact. A first approach would be to estimate direct impacts such as potential energy saved by individual settings, taking into account their default options and other strategies we discussed. However, this approach is misleading~\cite{ekchajzer_decision-making_2024} as emissions reduction induced by design decisions such as the ones we presented is notoriously difficult to compute and varies depending on the scope chosen. 
While some direct effects can be measured, there is no established method for indirect effects although they largely trump direct ones~\cite{pasek_world_2023}. 
Another challenge is that the most straightforward measures relate to energy consumption, which in practice obfuscates more complex evaluations of environmental impacts, whether it relates to mining, manufacturing, e-waste, and their effects on soil degradation, water consumption or pollution,~etc.

Aligned with Santarius, Tilman et al.'s \cite{santarius_digital_2022} approach to digital sufficiency, we invite further research for capturing a more holistic understanding of design choices' impact regarding e.g. hardware sufficiency or pressure on ICT infrastructure capacity. For example, whether settings can support extending the lifespan of devices by making the use of aging devices more ``bearable'' is something we only started exploring in our design workbook.

In this paper, our interests also lie in the potential of design to change perceptions. Design has traditionally played a role in domesticating innovation, facilitating the adoption of new products and services, and promoting consumption. Today, the general marketing discourse is still largely based on the idea that more is better (more storage space, more performance, faster network...). As we have seen in our analysis, options for more moderate uses are currently seen as a downgrade, a worse experience. Design choices, including visual ones, participate in shaping users expectations and perceptions and should be further explored. 

\subsection{The Systemic Elements of Settings}
In practice, the availability of settings, i.e., what can be controlled by end-users (e.g., video resolution, data flows) is defined by service providers. OS and applications defaults are also beyond end-users intervention. Collectively defining settings, their default options, or controlling internal factors not available as settings (e.g., the compression algorithm of a video) could have significantly more impact than controlling settings on an individual level.

Moreover, as we saw in our analysis, the amount of settings greatly depends on the business model of digital services and applications. YouTube does not provide an audio only option as this is something they currently sell; mainstream social media platforms do not let users delete content as they use and sell this data, whereas open source alternatives tend to offer more control. Looking at application settings (or their absence) offers a window into the strategic decisions of the organization producing these digital products and services\footnote{Zeynep Tufekci. 2023. Opinion | This Is Why Google Paid Billions for Apple to Change a Single Setting. The New York Times. Retrieved February 7, 2024 from \url{https://www.nytimes.com/2023/11/20/opinion/apple-google-privacy.html}}. While the design of settings is conditioned by the regulatory and economic environment in which applications are developed, studies like ours show that alternatives are possible and should be considered.

\subsection{Integrating User-Led With System-Led Ecological Customization}
Rather than opposing user-controlled customization and systemic changes, we think that these are two facets of change to be tackled together. While digital services providers should work on reducing energy consumption, data transfer and storage volumes, maximizing hardware lifetime, and managing use intensity, individuals should maintain some level of agency.
Indeed, only users can know when they need to watch a video in high quality mode or when only listening to it would be enough. Widdicks \& Pargman, in their discussion of moderate Internet use~\cite{widdicks_breaking_2019}, also argue for a balance between a user-controlled and a technology-controlled approach. 
Such balance could be used to facilitate transition to more moderate or ecological uses, or coming back to intensive use when needed. In our design workbook, our goal was to prioritize sustainability but this shift in priority can lead to obscuring the meaning of settings in terms of functionality. Future work could expand our design workbook and explore how to integrate a system-led ecological personalization that supports shifting priorities while letting people remain in control.

\subsection{Settings Beyond the individual}
We are aware of the risks of generating ``settings bloat''. Offering many settings options can be overwhelming, which can explain why they are sometimes underused in practice \cite{xu_hey_2015}. Designing and implementing more options might even make a given software heavier which could be counterproductive. Rather than suggesting an increase in the settings available, we argue for a shift on what settings should enable and how to communicate their effects. Frugal defaults are one element, but collective settings are another direction. 
In the workbook, we presented strategies to share settings or collectively vote for default options. These are only a sample of broader possibilities that should be further explored. In professional or educational contexts, settings are already defined by system administrators, whether to implement some organization policy or to enforce legal requirements (e.g., the right to disconnect). In households, Chetty et al. already demonstrated how bandwidth and home network parameters could be managed at a family level \cite{chetty_whos_2010}. In those collective settings, future work could study how settings could be defined, negotiated, and implemented. Future work could test these alternative and collectively defined settings in context; it would help further understand settings efficiency and the power dynamics at play.

\subsection{On the Risks of Settings Washing}
The argument we put forward should not minimize the risks of ``settings washing'', i.e., companies introducing environmental settings at the margins and being vocal about how they are letting end-users control their environmental impact. 
This could be compared to ``do-good practices'' such as emptying one’s inbox, i.e., practices that are widely advertised but have a very limited impact and, akin to greenwashing, divert from much more impactful decisions \cite{joshi_search_2015}. 
This is particularly relevant as Hazas et al. point out a rising proportion of data demand responding not to consumers' actions but to machine-to-machine communication~\cite{hazas_are_2016}, such as software updates, sensor data transferred to data-centers, data backups, load balancing, etc. Recent investments in data-centers are also largely driven by speculation around AI and hypothetical market capture, rather than genuine demand\footnote{Five reasons to question the frenzy behind the technology, Tej Parikh, The Financial Times, 2024-07-18, \url{https://www.ft.com/content/42bad56f-02cc-4b32-b9ac-1af5dbc7bc83}}.

Nonetheless, designing environmental settings remains a necessity. It participates in making environmental impacts more visible and contributes to supporting perspective shifts, especially by reminding of the material impacts of digital infrastructures.
Akin to recycling by individuals, which would still be needed in a world with much less plastic and disposable objects, settings would still be needed in a world with much more efficient and frugal digital systems.

\section{CONCLUSION}

In this paper, we were interested in understanding to what extent smartphone settings support or prevent people from adopting environmentally aware practices. Our analysis led us to \emph{identify and define 6 anti-patterns that explain how current smartphone settings designs do not support users in making ecological choices}. Based on these empirical results, we proposed eight design principles for environmental settings and tested their generative potential through the creation of alternative settings designs. 
Our design proposals pave the way for more collective deliberation around environmental settings. Today, settings are mostly available at the device level or at the IT management level, but we believe that there are opportunities for new collective setting practices in shared devices contexts such as schools and libraries.
Our work also calls for more research around understanding the impact of such actions, beyond net impacts and towards understanding indirect effects such as promoting longer device life.

\section{Acknowledgments}
We would like to thank all the workshop participants for their time and insights. We also thank reviewers for their constructive feedback. This work was partially supported by the ANR (Agence Nationale de la Recherche) grants: Suffisance Numérique - ANR-23-SSAI-0022 and PEPR Ensemble – ANR-22-EXEN-0006 (PEPR eNSEMBLE / PC5).


\bibliographystyle{ACM-Reference-Format}
\bibliography{CHI25}


\begin{thebibliography}{46}


\ifx \showCODEN    \undefined \def \showCODEN     #1{\unskip}     \fi
\ifx \showDOI      \undefined \def \showDOI       #1{#1}\fi
\ifx \showISBNx    \undefined \def \showISBNx     #1{\unskip}     \fi
\ifx \showISBNxiii \undefined \def \showISBNxiii  #1{\unskip}     \fi
\ifx \showISSN     \undefined \def \showISSN      #1{\unskip}     \fi
\ifx \showLCCN     \undefined \def \showLCCN      #1{\unskip}     \fi
\ifx \shownote     \undefined \def \shownote      #1{#1}          \fi
\ifx \showarticletitle \undefined \def \showarticletitle #1{#1}   \fi
\ifx \showURL      \undefined \def \showURL       {\relax}        \fi
\providecommand\bibfield[2]{#2}
\providecommand\bibinfo[2]{#2}
\providecommand\natexlab[1]{#1}
\providecommand\showeprint[2][]{arXiv:#2}

\bibitem[Bailly et~al\mbox{.}(2017)]%
        {bailly_visual_2017}
\bibfield{author}{\bibinfo{person}{Gilles Bailly}, \bibinfo{person}{{\'E}ric
  Lecolinet}, {and} \bibinfo{person}{Laurence Nigay}.}
  \bibinfo{year}{2017}\natexlab{}.
\newblock \showarticletitle{Visual {{Menu Techniques}}}.
\newblock \bibinfo{journal}{\emph{Comput. Surveys}} \bibinfo{volume}{49},
  \bibinfo{number}{4} (\bibinfo{year}{2017}), \bibinfo{pages}{60}.
\newblock
\urldef\tempurl%
\url{https://doi.org/10.1145/3002171}
\showDOI{\tempurl}


\bibitem[Bates et~al\mbox{.}(2015)]%
        {bates_exploring_2015}
\bibfield{author}{\bibinfo{person}{Oliver Bates}, \bibinfo{person}{Carolynne
  Lord}, \bibinfo{person}{Bran Knowles}, \bibinfo{person}{Adrian Friday},
  \bibinfo{person}{Adrian Clear}, {and} \bibinfo{person}{Mike Hazas}.}
  \bibinfo{year}{2015}\natexlab{}.
\newblock \showarticletitle{Exploring (Un)Sustainable Growth of Digital
  Technologies in the Home}. In \bibinfo{booktitle}{\emph{{{EnviroInfo}} and
  {{ICT}} for {{Sustainability}} 2015}}. \bibinfo{publisher}{Atlantis Press},
  \bibinfo{pages}{300--309}.
\newblock
\showISBNx{978-94-6252-092-9}
\showISSN{2352-538X}
\urldef\tempurl%
\url{https://doi.org/10.2991/ict4s-env-15.2015.34}
\showDOI{\tempurl}


\bibitem[Beignon(2021)]%
        {beignon_design_2021}
\bibfield{author}{\bibinfo{person}{Ana{\"e}lle Beignon}.}
  \bibinfo{year}{2021}\natexlab{}.
\newblock \bibinfo{booktitle}{\emph{Design for Obsolete Devices. :
  {{Exploring}} the Marginalization of Users of Obsolete Devices Regarding the
  {{Swedish}} Public Services' Digitalization.}}
\newblock \bibinfo{type}{Masters Thesis}. \bibinfo{institution}{{Malm{\"o}
  University, School of Arts and Communication (K3) / Malm{\"o} University,
  School of Arts and Communication (K3)}}. \bibinfo{pages}{58} pages.
\newblock


\bibitem[Blackwell({[n.\,d.]})]%
        {blackwell_reification_2006}
\bibfield{author}{\bibinfo{person}{Alan~F. Blackwell}.}
  \bibinfo{year}{[n.\,d.]}\natexlab{}.
\newblock \showarticletitle{The Reification of Metaphor as a Design Tool}.
\newblock  \bibinfo{volume}{13}, \bibinfo{number}{4}
  (\bibinfo{year}{[n.\,d.]}), \bibinfo{pages}{490--530}.
\newblock
\showISSN{1073-0516}
\urldef\tempurl%
\url{https://doi.org/10.1145/1188816.1188820}
\showDOI{\tempurl}


\bibitem[Blevis(2007)]%
        {blevis_sustainable_2007}
\bibfield{author}{\bibinfo{person}{Eli Blevis}.}
  \bibinfo{year}{2007}\natexlab{}.
\newblock \showarticletitle{Sustainable Interaction Design: Invention \&
  Disposal, Renewal \& Reuse}. In \bibinfo{booktitle}{\emph{Proceedings of the
  {{SIGCHI Conference}} on {{Human Factors}} in {{Computing Systems}}}}
  \emph{(\bibinfo{series}{{{CHI}} '07})}. \bibinfo{publisher}{Association for
  Computing Machinery}, \bibinfo{address}{San Jose, California, USA},
  \bibinfo{pages}{503--512}.
\newblock
\showISBNx{978-1-59593-593-9}
\urldef\tempurl%
\url{https://doi.org/10.1145/1240624.1240705}
\showDOI{\tempurl}


\bibitem[Bremer et~al\mbox{.}(2022)]%
        {bremer_have_2022}
\bibfield{author}{\bibinfo{person}{Christina Bremer}, \bibinfo{person}{Bran
  Knowles}, {and} \bibinfo{person}{Adrian Friday}.}
  \bibinfo{year}{2022}\natexlab{}.
\newblock \showarticletitle{Have {{We Taken On Too Much}}?: {{A Critical
  Review}} of the {{Sustainable HCI Landscape}}}. In
  \bibinfo{booktitle}{\emph{{{CHI Conference}} on {{Human Factors}} in
  {{Computing Systems}}}}. \bibinfo{publisher}{ACM}, \bibinfo{address}{New
  Orleans LA USA}, \bibinfo{pages}{1--11}.
\newblock
\showISBNx{978-1-4503-9157-3}
\urldef\tempurl%
\url{https://doi.org/10.1145/3491102.3517609}
\showDOI{\tempurl}


\bibitem[Brown et~al\mbox{.}(1998)]%
        {brown1998antipatterns}
\bibfield{author}{\bibinfo{person}{William~H Brown}, \bibinfo{person}{Raphael~C
  Malveau}, \bibinfo{person}{Hays W"~Skip" McCormick}, {and}
  \bibinfo{person}{Thomas~J Mowbray}.} \bibinfo{year}{1998}\natexlab{}.
\newblock \bibinfo{booktitle}{\emph{AntiPatterns: refactoring software,
  architectures, and projects in crisis}}.
\newblock \bibinfo{publisher}{John Wiley \& Sons, Inc.}
\newblock


\bibitem[Chetty et~al\mbox{.}(2010)]%
        {chetty_whos_2010}
\bibfield{author}{\bibinfo{person}{Marshini Chetty}, \bibinfo{person}{Richard
  Banks}, \bibinfo{person}{Richard Harper}, \bibinfo{person}{Tim Regan},
  \bibinfo{person}{Abigail Sellen}, \bibinfo{person}{Christos Gkantsidis},
  \bibinfo{person}{Thomas Karagiannis}, {and} \bibinfo{person}{Peter Key}.}
  \bibinfo{year}{2010}\natexlab{}.
\newblock \showarticletitle{Who's Hogging the Bandwidth: The Consequences of
  Revealing the Invisible in the Home}. In
  \bibinfo{booktitle}{\emph{Proceedings of the 28th International Conference on
  {{Human}} Factors in Computing Systems - {{CHI}} '10}}.
  \bibinfo{publisher}{ACM Press}, \bibinfo{address}{Atlanta, Georgia, USA},
  \bibinfo{pages}{659}.
\newblock
\showISBNx{978-1-60558-929-9}
\urldef\tempurl%
\url{https://doi.org/10.1145/1753326.1753423}
\showDOI{\tempurl}


\bibitem[Dourish(2010)]%
        {dourish_hci_2010}
\bibfield{author}{\bibinfo{person}{Paul Dourish}.}
  \bibinfo{year}{2010}\natexlab{}.
\newblock \showarticletitle{{{HCI}} and Environmental Sustainability: The
  Politics of Design and the Design of Politics}. In
  \bibinfo{booktitle}{\emph{Proceedings of the 8th {{ACM Conference}} on
  {{Designing Interactive Systems}}}} \emph{(\bibinfo{series}{{{DIS}} '10})}.
  \bibinfo{publisher}{Association for Computing Machinery},
  \bibinfo{address}{New York, NY, USA}, \bibinfo{pages}{1--10}.
\newblock
\showISBNx{978-1-4503-0103-9}
\urldef\tempurl%
\url{https://doi.org/10.1145/1858171.1858173}
\showDOI{\tempurl}


\bibitem[Ekchajzer et~al\mbox{.}(2024)]%
        {ekchajzer_decision-making_2024}
\bibfield{author}{\bibinfo{person}{David Ekchajzer}, \bibinfo{person}{Laetitia
  Bornes}, \bibinfo{person}{Jacques Combaz}, \bibinfo{person}{Catherine
  Letondal}, {and} \bibinfo{person}{Rob Vingerhoeds}.}
  \bibinfo{year}{2024}\natexlab{}.
\newblock \showarticletitle{Decision-{{Making Under Environmental Complexity}}:
  {{The Need}} for {{Moving}} from {{Avoided Impacts}} of {{ICT Solutions}} to
  {{Systems Thinking Approaches}}}. In \bibinfo{booktitle}{\emph{2024 10th
  {{International Conference}} on {{ICT}} for {{Sustainability}} ({{ICT4S}})}}.
  \bibinfo{pages}{29--40}.
\newblock
\urldef\tempurl%
\url{https://doi.org/10.1109/ICT4S64576.2024.00013}
\showDOI{\tempurl}


\bibitem[Freitag et~al\mbox{.}(2021)]%
        {freitag_real_2021}
\bibfield{author}{\bibinfo{person}{Charlotte Freitag}, \bibinfo{person}{Mike
  {Berners-Lee}}, \bibinfo{person}{Kelly Widdicks}, \bibinfo{person}{Bran
  Knowles}, \bibinfo{person}{Gordon~S. Blair}, {and} \bibinfo{person}{Adrian
  Friday}.} \bibinfo{year}{2021}\natexlab{}.
\newblock \showarticletitle{The Real Climate and Transformative Impact of
  {{ICT}}: {{A}} Critique of Estimates, Trends, and Regulations}.
\newblock \bibinfo{journal}{\emph{Patterns}} \bibinfo{volume}{2},
  \bibinfo{number}{9} (\bibinfo{date}{Sept.} \bibinfo{year}{2021}),
  \bibinfo{pages}{100340}.
\newblock
\showISSN{2666-3899}
\urldef\tempurl%
\url{https://doi.org/10.1016/j.patter.2021.100340}
\showDOI{\tempurl}


\bibitem[Gaver(2011)]%
        {gaver_making_2011}
\bibfield{author}{\bibinfo{person}{William Gaver}.}
  \bibinfo{year}{2011}\natexlab{}.
\newblock \showarticletitle{Making Spaces: How Design Workbooks Work}. In
  \bibinfo{booktitle}{\emph{Proceedings of the {{SIGCHI Conference}} on {{Human
  Factors}} in {{Computing Systems}}}} \emph{(\bibinfo{series}{{{CHI}} '11})}.
  \bibinfo{publisher}{Association for Computing Machinery},
  \bibinfo{address}{New York, NY, USA}, \bibinfo{pages}{1551--1560}.
\newblock
\showISBNx{978-1-4503-0228-9}
\urldef\tempurl%
\url{https://doi.org/10.1145/1978942.1979169}
\showDOI{\tempurl}


\bibitem[Hasinoff and Bivens(2021)]%
        {hasinoff_feature_2021}
\bibfield{author}{\bibinfo{person}{Amy Hasinoff} {and} \bibinfo{person}{Rena
  Bivens}.} \bibinfo{year}{2021}\natexlab{}.
\newblock \showarticletitle{Feature {{Analysis}}: {{A Method}} for
  {{Analyzing}} the {{Role}} of {{Ideology}} in {{App Design}}}.
\newblock \bibinfo{journal}{\emph{Journal of Digital Social Research}}
  \bibinfo{volume}{3}, \bibinfo{number}{2} (\bibinfo{date}{Sept.}
  \bibinfo{year}{2021}), \bibinfo{pages}{89--113}.
\newblock
\showISSN{2003-1998}
\urldef\tempurl%
\url{https://doi.org/10.33621/jdsr.v3i2.56}
\showDOI{\tempurl}


\bibitem[Hazas et~al\mbox{.}(2016)]%
        {hazas_are_2016}
\bibfield{author}{\bibinfo{person}{Mike Hazas}, \bibinfo{person}{Janine
  Morley}, \bibinfo{person}{Oliver Bates}, {and} \bibinfo{person}{Adrian
  Friday}.} \bibinfo{year}{2016}\natexlab{}.
\newblock \showarticletitle{Are There Limits to Growth in Data Traffic? On Time
  Use, Data Generation and Speed}. In \bibinfo{booktitle}{\emph{Proceedings of
  the {{Second Workshop}} on {{Computing}} within {{Limits}}}}
  \emph{(\bibinfo{series}{{{LIMITS}} '16})}. \bibinfo{publisher}{Association
  for Computing Machinery}, \bibinfo{address}{New York, NY, USA},
  \bibinfo{pages}{1--5}.
\newblock
\showISBNx{978-1-4503-4260-5}
\urldef\tempurl%
\url{https://doi.org/10.1145/2926676.2926690}
\showDOI{\tempurl}


\bibitem[Hill et~al\mbox{.}(2020)]%
        {hill_mapping_2020}
\bibfield{author}{\bibinfo{person}{Joshua Hill}, \bibinfo{person}{Kelly
  Widdicks}, {and} \bibinfo{person}{Mike Hazas}.}
  \bibinfo{year}{2020}\natexlab{}.
\newblock \showarticletitle{Mapping the {{Scope}} of {{Software Interventions}}
  for {{Moderate Internet Use}} on {{Mobile Devices}}}. In
  \bibinfo{booktitle}{\emph{Proceedings of the 7th {{International Conference}}
  on {{ICT}} for {{Sustainability}}}} \emph{(\bibinfo{series}{{{ICT4S2020}}})}.
  \bibinfo{publisher}{Association for Computing Machinery},
  \bibinfo{address}{New York, NY, USA}, \bibinfo{pages}{204--212}.
\newblock
\showISBNx{978-1-4503-7595-5}
\urldef\tempurl%
\url{https://doi.org/10.1145/3401335.3401361}
\showDOI{\tempurl}


\bibitem[Huang et~al\mbox{.}(2012)]%
        {huang_close_2012}
\bibfield{author}{\bibinfo{person}{Junxian Huang}, \bibinfo{person}{Feng Qian},
  \bibinfo{person}{Alexandre Gerber}, \bibinfo{person}{Z.~Morley Mao},
  \bibinfo{person}{Subhabrata Sen}, {and} \bibinfo{person}{Oliver Spatscheck}.}
  \bibinfo{year}{2012}\natexlab{}.
\newblock \showarticletitle{A Close Examination of Performance and Power
  Characteristics of {{4G LTE}} Networks}. In
  \bibinfo{booktitle}{\emph{Proceedings of the 10th International Conference on
  {{Mobile}} Systems, Applications, and Services}}
  \emph{(\bibinfo{series}{{{MobiSys}} '12})}. \bibinfo{publisher}{Association
  for Computing Machinery}, \bibinfo{address}{New York, NY, USA},
  \bibinfo{pages}{225--238}.
\newblock
\showISBNx{978-1-4503-1301-8}
\urldef\tempurl%
\url{https://doi.org/10.1145/2307636.2307658}
\showDOI{\tempurl}


\bibitem[Joshi and Pargman(2015)]%
        {joshi_search_2015}
\bibfield{author}{\bibinfo{person}{Somya Joshi} {and}
  \bibinfo{person}{Teresa~Cerratto Pargman}.} \bibinfo{year}{2015}\natexlab{}.
\newblock \showarticletitle{In {{Search}} of {{Fairness}}: {{Critical Design
  Alternatives}} for {{Sustainability}}}.
\newblock \bibinfo{journal}{\emph{Aarhus Series on Human Centered Computing}}
  \bibinfo{volume}{1}, \bibinfo{number}{1} (\bibinfo{date}{Oct.}
  \bibinfo{year}{2015}), \bibinfo{pages}{4--4}.
\newblock
\showISSN{2445-7221}
\urldef\tempurl%
\url{https://doi.org/10.7146/aahcc.v1i1.21301}
\showDOI{\tempurl}


\bibitem[Kara et~al\mbox{.}(2019)]%
        {kara_comparison_2019}
\bibfield{author}{\bibinfo{person}{Peter~A. Kara}, \bibinfo{person}{Werner
  Robitza}, \bibinfo{person}{Nikolett Pinter}, \bibinfo{person}{Maria~G.
  Martini}, \bibinfo{person}{Alexander Raake}, {and} \bibinfo{person}{Aniko
  Simon}.} \bibinfo{year}{2019}\natexlab{}.
\newblock \showarticletitle{Comparison of {{HD}} and {{UHD}} Video Quality with
  and without the Influence of the Labeling Effect}.
\newblock \bibinfo{journal}{\emph{Quality and User Experience}}
  \bibinfo{volume}{4}, \bibinfo{number}{1} (\bibinfo{date}{Sept.}
  \bibinfo{year}{2019}), \bibinfo{pages}{4}.
\newblock
\showISSN{2366-0147}
\urldef\tempurl%
\url{https://doi.org/10.1007/s41233-019-0027-3}
\showDOI{\tempurl}


\bibitem[Kurp(2008)]%
        {kurp_green_2008}
\bibfield{author}{\bibinfo{person}{Patrick Kurp}.}
  \bibinfo{year}{2008}\natexlab{}.
\newblock \showarticletitle{Green Computing}.
\newblock \bibinfo{journal}{\emph{Commun. ACM}} \bibinfo{volume}{51},
  \bibinfo{number}{10} (\bibinfo{date}{Oct.} \bibinfo{year}{2008}),
  \bibinfo{pages}{11--13}.
\newblock
\showISSN{0001-0782}
\urldef\tempurl%
\url{https://doi.org/10.1145/1400181.1400186}
\showDOI{\tempurl}


\bibitem[Lef{\`e}vre and Orgerie(2010)]%
        {lefevre_designing_2010}
\bibfield{author}{\bibinfo{person}{Laurent Lef{\`e}vre} {and}
  \bibinfo{person}{Anne-C{\'e}cile Orgerie}.} \bibinfo{year}{2010}\natexlab{}.
\newblock \showarticletitle{Designing and Evaluating an Energy Efficient
  {{Cloud}}}.
\newblock \bibinfo{journal}{\emph{The Journal of Supercomputing}}
  \bibinfo{volume}{51}, \bibinfo{number}{3} (\bibinfo{date}{March}
  \bibinfo{year}{2010}), \bibinfo{pages}{352--373}.
\newblock
\showISSN{1573-0484}
\urldef\tempurl%
\url{https://doi.org/10.1007/s11227-010-0414-2}
\showDOI{\tempurl}


\bibitem[Lord et~al\mbox{.}(2015)]%
        {lord_demand_2015}
\bibfield{author}{\bibinfo{person}{Carolynne Lord}, \bibinfo{person}{Mike
  Hazas}, \bibinfo{person}{Adrian~K. Clear}, \bibinfo{person}{Oliver Bates},
  \bibinfo{person}{Rosalind Whittam}, \bibinfo{person}{Janine Morley}, {and}
  \bibinfo{person}{Adrian Friday}.} \bibinfo{year}{2015}\natexlab{}.
\newblock \showarticletitle{Demand in {{My Pocket}}: {{Mobile Devices}} and the
  {{Data Connectivity Marshalled}} in {{Support}} of {{Everyday Practice}}}. In
  \bibinfo{booktitle}{\emph{Proceedings of the 33rd {{Annual ACM Conference}}
  on {{Human Factors}} in {{Computing Systems}}}}. \bibinfo{publisher}{ACM},
  \bibinfo{address}{Seoul Republic of Korea}, \bibinfo{pages}{2729--2738}.
\newblock
\showISBNx{978-1-4503-3145-6}
\urldef\tempurl%
\url{https://doi.org/10.1145/2702123.2702162}
\showDOI{\tempurl}


\bibitem[Lorusso(2021)]%
        {lorusso_user_2021}
\bibfield{author}{\bibinfo{person}{Silvio Lorusso}.}
  \bibinfo{year}{2021}\natexlab{}.
\newblock \bibinfo{title}{The {{User Condition}}}.
\newblock
\newblock


\bibitem[Mackay(1991)]%
        {mackay_triggers_1991}
\bibfield{author}{\bibinfo{person}{Wendy~E. Mackay}.}
  \bibinfo{year}{1991}\natexlab{}.
\newblock \showarticletitle{Triggers and Barriers to Customizing Software}. In
  \bibinfo{booktitle}{\emph{Proceedings of the {{SIGCHI Conference}} on {{Human
  Factors}} in {{Computing Systems}}}} \emph{(\bibinfo{series}{{{CHI}} '91})}.
  \bibinfo{publisher}{Association for Computing Machinery},
  \bibinfo{address}{New York, NY, USA}, \bibinfo{pages}{153--160}.
\newblock
\showISBNx{978-0-89791-383-6}
\urldef\tempurl%
\url{https://doi.org/10.1145/108844.108867}
\showDOI{\tempurl}


\bibitem[Maxwell(2010)]%
        {maxwell_using_2010}
\bibfield{author}{\bibinfo{person}{Joseph~A. Maxwell}.}
  \bibinfo{year}{2010}\natexlab{}.
\newblock \showarticletitle{Using {{Numbers}} in {{Qualitative Research}}}.
\newblock \bibinfo{journal}{\emph{Qualitative Inquiry}} \bibinfo{volume}{16},
  \bibinfo{number}{6} (\bibinfo{date}{July} \bibinfo{year}{2010}),
  \bibinfo{pages}{475--482}.
\newblock
\showISSN{1077-8004}
\urldef\tempurl%
\url{https://doi.org/10.1177/1077800410364740}
\showDOI{\tempurl}


\bibitem[McGrenere et~al\mbox{.}(2002)]%
        {mcgrenere_evaluation_2002}
\bibfield{author}{\bibinfo{person}{Joanna McGrenere},
  \bibinfo{person}{Ronald~M. Baecker}, {and} \bibinfo{person}{Kellogg~S.
  Booth}.} \bibinfo{year}{2002}\natexlab{}.
\newblock \showarticletitle{An Evaluation of a Multiple Interface Design
  Solution for Bloated Software}. In \bibinfo{booktitle}{\emph{Proceedings of
  the {{SIGCHI Conference}} on {{Human Factors}} in {{Computing Systems}}}}
  \emph{(\bibinfo{series}{{{CHI}} '02})}. \bibinfo{publisher}{Association for
  Computing Machinery}, \bibinfo{address}{New York, NY, USA},
  \bibinfo{pages}{164--170}.
\newblock
\showISBNx{978-1-58113-453-7}
\urldef\tempurl%
\url{https://doi.org/10.1145/503376.503406}
\showDOI{\tempurl}


\bibitem[McGrenere and Moore(2000)]%
        {mcgrenere_are_2000}
\bibfield{author}{\bibinfo{person}{Joanna McGrenere} {and}
  \bibinfo{person}{Gale Moore}.} \bibinfo{year}{2000}\natexlab{}.
\newblock \showarticletitle{Are {{We All}} in the {{Same}} "{{Bloat}}"?}
\newblock \bibinfo{journal}{\emph{Proceedings of Graphics Interface 2000}}
  \bibinfo{volume}{Montr{\'e}al} (\bibinfo{year}{2000}), \bibinfo{pages}{10
  pages, 143.23 KB}.
\newblock
\showISBNx{9780969533894}
\showISSN{0713-5424}
\urldef\tempurl%
\url{https://doi.org/10.20380/GI2000.25}
\showDOI{\tempurl}


\bibitem[Monge~Roffarello and De~Russis(2022)]%
        {monge_roffarello_towards_2022}
\bibfield{author}{\bibinfo{person}{Alberto Monge~Roffarello} {and}
  \bibinfo{person}{Luigi De~Russis}.} \bibinfo{year}{2022}\natexlab{}.
\newblock \showarticletitle{Towards {{Understanding}} the {{Dark Patterns That
  Steal Our Attention}}}. In \bibinfo{booktitle}{\emph{Extended {{Abstracts}}
  of the 2022 {{CHI Conference}} on {{Human Factors}} in {{Computing
  Systems}}}} \emph{(\bibinfo{series}{{{CHI EA}} '22})}.
  \bibinfo{publisher}{Association for Computing Machinery},
  \bibinfo{address}{New York, NY, USA}, \bibinfo{pages}{1--7}.
\newblock
\showISBNx{978-1-4503-9156-6}
\urldef\tempurl%
\url{https://doi.org/10.1145/3491101.3519829}
\showDOI{\tempurl}


\bibitem[Mosesso et~al\mbox{.}(2023)]%
        {mosesso_obsolescence_2023}
\bibfield{author}{\bibinfo{person}{L{\'e}a Mosesso}, \bibinfo{person}{Nolwenn
  Maudet}, \bibinfo{person}{Edlira Nano}, \bibinfo{person}{Thomas Thibault},
  {and} \bibinfo{person}{Aur{\'e}lien Tabard}.}
  \bibinfo{year}{2023}\natexlab{}.
\newblock \showarticletitle{Obsolescence {{Paths}}: Living with Aging Devices}.
  In \bibinfo{booktitle}{\emph{{{ICT4S}} 2023 - {{International Conference}} on
  {{Information}} and {{Communications Technology}} for {{Sustainability}}}}.
\newblock


\bibitem[Noureddine et~al\mbox{.}(2012)]%
        {noureddine_preliminary_2012}
\bibfield{author}{\bibinfo{person}{Adel Noureddine}, \bibinfo{person}{Aurelien
  Bourdon}, \bibinfo{person}{Romain Rouvoy}, {and} \bibinfo{person}{Lionel
  Seinturier}.} \bibinfo{year}{2012}\natexlab{}.
\newblock \showarticletitle{A Preliminary Study of the Impact of Software
  Engineering on {{GreenIT}}}. In \bibinfo{booktitle}{\emph{2012 {{First
  International Workshop}} on {{Green}} and {{Sustainable Software}}
  ({{GREENS}})}}. \bibinfo{pages}{21--27}.
\newblock
\urldef\tempurl%
\url{https://doi.org/10.1109/GREENS.2012.6224251}
\showDOI{\tempurl}


\bibitem[Pasek et~al\mbox{.}(2023)]%
        {pasek_world_2023}
\bibfield{author}{\bibinfo{person}{Anne Pasek}, \bibinfo{person}{Hunter
  Vaughan}, {and} \bibinfo{person}{Nicole Starosielski}.}
  \bibinfo{year}{2023}\natexlab{}.
\newblock \showarticletitle{The World Wide Web of Carbon: {{Toward}} a
  Relational Footprinting of Information and Communications Technology's
  Climate Impacts}.
\newblock \bibinfo{journal}{\emph{Big Data \& Society}} \bibinfo{volume}{10},
  \bibinfo{number}{1} (\bibinfo{date}{Jan.} \bibinfo{year}{2023}),
  \bibinfo{pages}{205395172311589}.
\newblock
\showISSN{2053-9517, 2053-9517}
\urldef\tempurl%
\url{https://doi.org/10.1177/20539517231158994}
\showDOI{\tempurl}


\bibitem[Preist et~al\mbox{.}(2016)]%
        {preist_understanding_2016}
\bibfield{author}{\bibinfo{person}{Chris Preist}, \bibinfo{person}{Daniel
  Schien}, {and} \bibinfo{person}{Eli Blevis}.}
  \bibinfo{year}{2016}\natexlab{}.
\newblock \showarticletitle{Understanding and {{Mitigating}} the {{Effects}} of
  {{Device}} and {{Cloud Service Design Decisions}} on the {{Environmental
  Footprint}} of {{Digital Infrastructure}}}. In
  \bibinfo{booktitle}{\emph{Proceedings of the 2016 {{CHI Conference}} on
  {{Human Factors}} in {{Computing Systems}}}}. \bibinfo{publisher}{ACM},
  \bibinfo{address}{San Jose California USA}, \bibinfo{pages}{1324--1337}.
\newblock
\showISBNx{978-1-4503-3362-7}
\urldef\tempurl%
\url{https://doi.org/10.1145/2858036.2858378}
\showDOI{\tempurl}


\bibitem[Preist et~al\mbox{.}(2019)]%
        {preist_evaluating_2019}
\bibfield{author}{\bibinfo{person}{Chris Preist}, \bibinfo{person}{Daniel
  Schien}, {and} \bibinfo{person}{Paul Shabajee}.}
  \bibinfo{year}{2019}\natexlab{}.
\newblock \showarticletitle{Evaluating {{Sustainable Interaction Design}} of
  {{Digital Services}}: {{The Case}} of {{YouTube}}}. In
  \bibinfo{booktitle}{\emph{Proceedings of the 2019 {{CHI Conference}} on
  {{Human Factors}} in {{Computing Systems}}}}. \bibinfo{publisher}{ACM},
  \bibinfo{address}{Glasgow Scotland Uk}, \bibinfo{pages}{1--12}.
\newblock
\showISBNx{978-1-4503-5970-2}
\urldef\tempurl%
\url{https://doi.org/10.1145/3290605.3300627}
\showDOI{\tempurl}


\bibitem[Santarius et~al\mbox{.}(2022)]%
        {santarius_digital_2022}
\bibfield{author}{\bibinfo{person}{Tilman Santarius}, \bibinfo{person}{Jan
  C.~T. Bieser}, \bibinfo{person}{Vivian Frick}, \bibinfo{person}{Mattias
  H{\"o}jer}, \bibinfo{person}{Maike Gossen}, \bibinfo{person}{Lorenz~M.
  Hilty}, \bibinfo{person}{Eva Kern}, \bibinfo{person}{Johanna Pohl},
  \bibinfo{person}{Friederike Rohde}, {and} \bibinfo{person}{Steffen Lange}.}
  \bibinfo{year}{2022}\natexlab{}.
\newblock \showarticletitle{Digital Sufficiency: Conceptual Considerations for
  {{ICTs}} on a Finite Planet}.
\newblock \bibinfo{journal}{\emph{Annals of Telecommunications}}
  (\bibinfo{date}{May} \bibinfo{year}{2022}).
\newblock
\showISSN{1958-9395}
\urldef\tempurl%
\url{https://doi.org/10.1007/s12243-022-00914-x}
\showDOI{\tempurl}


\bibitem[Schaub et~al\mbox{.}(2014)]%
        {schaub_broken_2014}
\bibfield{author}{\bibinfo{person}{Florian Schaub}, \bibinfo{person}{Julian
  Seifert}, \bibinfo{person}{Frank Honold}, \bibinfo{person}{Michael
  M{\"u}ller}, \bibinfo{person}{Enrico Rukzio}, {and} \bibinfo{person}{Michael
  Weber}.} \bibinfo{year}{2014}\natexlab{}.
\newblock \showarticletitle{Broken Display = Broken Interface': The Impact of
  Display Damage on Smartphone Interaction}. In
  \bibinfo{booktitle}{\emph{Proceedings of the {{SIGCHI Conference}} on {{Human
  Factors}} in {{Computing Systems}}}} \emph{(\bibinfo{series}{{{CHI}} '14})}.
  \bibinfo{publisher}{Association for Computing Machinery},
  \bibinfo{address}{New York, NY, USA}, \bibinfo{pages}{2337--2346}.
\newblock
\showISBNx{978-1-4503-2473-1}
\urldef\tempurl%
\url{https://doi.org/10.1145/2556288.2557067}
\showDOI{\tempurl}


\bibitem[Sundar and Marathe(2010)]%
        {sundar_personalization_2010}
\bibfield{author}{\bibinfo{person}{S.~Shyam Sundar} {and}
  \bibinfo{person}{Sampada~S. Marathe}.} \bibinfo{year}{2010}\natexlab{}.
\newblock \showarticletitle{Personalization versus {{Customization}}: The
  {{Importance}} of {{Agency}}, {{Privacy}}, and {{Power Usage}}}.
\newblock \bibinfo{journal}{\emph{Human Communication Research}}
  \bibinfo{volume}{36}, \bibinfo{number}{3} (\bibinfo{date}{July}
  \bibinfo{year}{2010}), \bibinfo{pages}{298--322}.
\newblock
\showISSN{0360-3989}
\urldef\tempurl%
\url{https://doi.org/10.1111/j.1468-2958.2010.01377.x}
\showDOI{\tempurl}


\bibitem[Suski et~al\mbox{.}(2020)]%
        {suski_all_2020}
\bibfield{author}{\bibinfo{person}{Paul Suski}, \bibinfo{person}{Johanna Pohl},
  {and} \bibinfo{person}{Vivian Frick}.} \bibinfo{year}{2020}\natexlab{}.
\newblock \showarticletitle{All You Can Stream: {{Investigating}} the Role of
  User Behavior for Greenhouse Gas Intensity of Video Streaming}. In
  \bibinfo{booktitle}{\emph{Proceedings of the 7th {{International Conference}}
  on {{ICT}} for {{Sustainability}}}}. \bibinfo{publisher}{ACM},
  \bibinfo{address}{Bristol United Kingdom}, \bibinfo{pages}{128--138}.
\newblock
\showISBNx{978-1-4503-7595-5}
\urldef\tempurl%
\url{https://doi.org/10.1145/3401335.3401709}
\showDOI{\tempurl}


\bibitem[Tchernavskij(2017)]%
        {tchernavskij_decomposing_2017}
\bibfield{author}{\bibinfo{person}{Philip Tchernavskij}.}
  \bibinfo{year}{2017}\natexlab{}.
\newblock \showarticletitle{Decomposing {{Interactive Systems}}}. In
  \bibinfo{booktitle}{\emph{{{CHI}} 2017 - {{Tools}} Workshop at
  {{CHI}}'2017}}. \bibinfo{address}{Denver, United States}, \bibinfo{pages}{4}.
\newblock


\bibitem[Vinet et~al\mbox{.}(2023)]%
        {vinet_everyday_2023}
\bibfield{author}{\bibinfo{person}{Louis Vinet}, \bibinfo{person}{Nolwenn
  Maudet}, {and} \bibinfo{person}{Aur{\'e}lien Tabard}.}
  \bibinfo{year}{2023}\natexlab{}.
\newblock \showarticletitle{The {{Everyday Experience}} of {{Connectivity
  Limits}} -- {{Insights}} from {{French Students}} during the {{Covid
  Pandemic}}}. In \bibinfo{booktitle}{\emph{Proceedings of the 34th
  {{Conference}} on l'{{Interaction Humain-Machine}}}}
  \emph{(\bibinfo{series}{{{IHM}} '23})}. \bibinfo{publisher}{Association for
  Computing Machinery}, \bibinfo{address}{New York, NY, USA},
  \bibinfo{pages}{1--11}.
\newblock
\showISBNx{978-1-4503-9824-4}
\urldef\tempurl%
\url{https://doi.org/10.1145/3583961.3583975}
\showDOI{\tempurl}


\bibitem[Vitale et~al\mbox{.}(2020)]%
        {vitale_data_2020}
\bibfield{author}{\bibinfo{person}{Francesco Vitale}, \bibinfo{person}{Janet
  Chen}, \bibinfo{person}{William Odom}, {and} \bibinfo{person}{Joanna
  McGrenere}.} \bibinfo{year}{2020}\natexlab{}.
\newblock \showarticletitle{Data {{Dashboard}}: {{Exploring Centralization}}
  and {{Customization}} in {{Personal Data Curation}}}. In
  \bibinfo{booktitle}{\emph{Proceedings of the 2020 {{ACM Designing Interactive
  Systems Conference}}}} \emph{(\bibinfo{series}{{{DIS}} '20})}.
  \bibinfo{publisher}{Association for Computing Machinery},
  \bibinfo{address}{New York, NY, USA}, \bibinfo{pages}{311--326}.
\newblock
\showISBNx{978-1-4503-6974-9}
\urldef\tempurl%
\url{https://doi.org/10.1145/3357236.3395457}
\showDOI{\tempurl}


\bibitem[Widdicks et~al\mbox{.}(2017)]%
        {widdicks_demand_2017}
\bibfield{author}{\bibinfo{person}{Kelly Widdicks}, \bibinfo{person}{Oliver
  Bates}, \bibinfo{person}{Mike Hazas}, \bibinfo{person}{Adrian Friday}, {and}
  \bibinfo{person}{Alastair~R. Beresford}.} \bibinfo{year}{2017}\natexlab{}.
\newblock \showarticletitle{Demand {{Around}} the {{Clock}}: {{Time Use}} and
  {{Data Demand}} of {{Mobile Devices}} in {{Everyday Life}}}. In
  \bibinfo{booktitle}{\emph{Proceedings of the 2017 {{CHI Conference}} on
  {{Human Factors}} in {{Computing Systems}}}} \emph{(\bibinfo{series}{{{CHI}}
  '17})}. \bibinfo{publisher}{Association for Computing Machinery},
  \bibinfo{address}{New York, NY, USA}, \bibinfo{pages}{5361--5372}.
\newblock
\showISBNx{978-1-4503-4655-9}
\urldef\tempurl%
\url{https://doi.org/10.1145/3025453.3025730}
\showDOI{\tempurl}


\bibitem[Widdicks et~al\mbox{.}(2019)]%
        {widdicks_streaming_2019}
\bibfield{author}{\bibinfo{person}{Kelly Widdicks}, \bibinfo{person}{Mike
  Hazas}, \bibinfo{person}{Oliver Bates}, {and} \bibinfo{person}{Adrian
  Friday}.} \bibinfo{year}{2019}\natexlab{}.
\newblock \showarticletitle{Streaming, {{Multi-Screens}} and {{YouTube}}: {{The
  New}} ({{Unsustainable}}) {{Ways}} of {{Watching}} in the {{Home}}}. In
  \bibinfo{booktitle}{\emph{Proceedings of the 2019 {{CHI Conference}} on
  {{Human Factors}} in {{Computing Systems}}}} \emph{(\bibinfo{series}{{{CHI}}
  '19})}. \bibinfo{publisher}{Association for Computing Machinery},
  \bibinfo{address}{New York, NY, USA}, \bibinfo{pages}{1--13}.
\newblock
\showISBNx{978-1-4503-5970-2}
\urldef\tempurl%
\url{https://doi.org/10.1145/3290605.3300696}
\showDOI{\tempurl}


\bibitem[Widdicks and Pargman(2019)]%
        {widdicks_breaking_2019}
\bibfield{author}{\bibinfo{person}{K. Widdicks} {and} \bibinfo{person}{Daniel
  Pargman}.} \bibinfo{year}{2019}\natexlab{}.
\newblock \showarticletitle{Breaking the Cornucopian Paradigm : {{Towards}}
  Moderate Internet Use in Everyday Life}. In \bibinfo{booktitle}{\emph{5th
  {{Workshop}} on {{Computing}} within {{Limits}}, {{LIMITS}} 2019,
  {{Lappeenranta}}, {{Finland}}, 10-11 {{June}} 2019}}.
  \bibinfo{publisher}{Association for Computing Machinery}.
\newblock


\bibitem[Widdicks et~al\mbox{.}(2020)]%
        {widdicks2020backfiring}
\bibfield{author}{\bibinfo{person}{Kelly Widdicks}, \bibinfo{person}{Daniel
  Pargman}, {and} \bibinfo{person}{Staffan Bjork}.}
  \bibinfo{year}{2020}\natexlab{}.
\newblock \showarticletitle{Backfiring and favouring: how design processes in
  HCI lead to anti-patterns and repentant designers}. In
  \bibinfo{booktitle}{\emph{Proceedings of the 11th Nordic Conference on
  Human-Computer Interaction: Shaping Experiences, Shaping Society}}.
  \bibinfo{pages}{1--12}.
\newblock


\bibitem[Widdicks et~al\mbox{.}(2022)]%
        {widdicks_escaping_2022}
\bibfield{author}{\bibinfo{person}{Kelly Widdicks}, \bibinfo{person}{Christian
  Remy}, \bibinfo{person}{Oliver Bates}, \bibinfo{person}{Adrian Friday}, {and}
  \bibinfo{person}{Mike Hazas}.} \bibinfo{year}{2022}\natexlab{}.
\newblock \showarticletitle{Escaping Unsustainable Digital Interactions:
  {{Toward}} ``More Meaningful'' and ``Moderate'' Online Experiences}.
\newblock \bibinfo{journal}{\emph{International Journal of Human-Computer
  Studies}}  \bibinfo{volume}{165} (\bibinfo{date}{Sept.}
  \bibinfo{year}{2022}), \bibinfo{pages}{102853}.
\newblock
\showISSN{1071-5819}
\urldef\tempurl%
\url{https://doi.org/10.1016/j.ijhcs.2022.102853}
\showDOI{\tempurl}


\bibitem[Widdicks et~al\mbox{.}(2018)]%
        {widdicks_undesigning_2018}
\bibfield{author}{\bibinfo{person}{Kelly Widdicks}, \bibinfo{person}{Tina
  Ringenson}, \bibinfo{person}{Daniel Pargman}, \bibinfo{person}{Vishnupriya
  Kuppusamy}, {and} \bibinfo{person}{Patricia Lago}.}
  \bibinfo{year}{2018}\natexlab{}.
\newblock \showarticletitle{Undesigning the {{Internet}}: {{An}} Exploratory
  Study of Reducing Everyday {{Internet}} Connectivity}. In
  \bibinfo{booktitle}{\emph{{{ICT4S2018}}. 5th {{International Conference}} on
  {{Information}} and {{Communication Technology}} for {{Sustainability}}}}.
  \bibinfo{pages}{384--369}.
\newblock
\urldef\tempurl%
\url{https://doi.org/10.29007/s221}
\showDOI{\tempurl}


\bibitem[Xu et~al\mbox{.}(2015)]%
        {xu_hey_2015}
\bibfield{author}{\bibinfo{person}{Tianyin Xu}, \bibinfo{person}{Long Jin},
  \bibinfo{person}{Xuepeng Fan}, \bibinfo{person}{Yuanyuan Zhou},
  \bibinfo{person}{Shankar Pasupathy}, {and} \bibinfo{person}{Rukma
  Talwadker}.} \bibinfo{year}{2015}\natexlab{}.
\newblock \showarticletitle{Hey, You Have given Me Too Many Knobs!:
  Understanding and Dealing with over-Designed Configuration in System
  Software}. In \bibinfo{booktitle}{\emph{Proceedings of the 2015 10th {{Joint
  Meeting}} on {{Foundations}} of {{Software Engineering}}}}
  \emph{(\bibinfo{series}{{{ESEC}}/{{FSE}} 2015})}.
  \bibinfo{publisher}{Association for Computing Machinery},
  \bibinfo{address}{New York, NY, USA}, \bibinfo{pages}{307--319}.
\newblock
\showISBNx{978-1-4503-3675-8}
\urldef\tempurl%
\url{https://doi.org/10.1145/2786805.2786852}
\showDOI{\tempurl}


\end{thebibliography}

\end{document}